\documentclass[12pt]{article}           % LaTeX 2e
\usepackage{hyperref}
\usepackage[pdftex]{graphicx}
\usepackage{amsmath,amssymb}
\usepackage{mathabx}
\usepackage{xcolor}
\usepackage{cite}
\usepackage{physics}
\newcommand{\be}{\begin{equation}}
\newcommand{\ee}{\end{equation}}
\newcommand{\bea}{\begin{eqnarray}}
\newcommand{\eea}{\end{eqnarray}}
\newcommand*\DAlambert{\mathop{}\!\mathbin\Box}

\newcommand{\gn}{\ensuremath{G_{\mbox{\scriptsize N}}}}

\def\eps{\ensuremath{\epsilon}}

\def\p{\partial}

\def\tr{\mathop{\rm Tr}}

\def\expec#1{\langle #1 \rangle}
\def\ket#1{| #1 \rangle}
\def\bra#1{\langle  #1 |}

\newcommand{\cC}{{\mathcal{C}}}
\newcommand{\cD}{{\mathcal{D}}}
\newcommand{\cE}{{\mathcal{E}}}

\newcommand{\cO}{{\mathcal{O}}}
\newcommand{\cP}{{\mathcal{P}}}
\newcommand{\cR}{{\mathcal{R}}}

\def\iir{\ensuremath{I_{\mbox{\tiny IR}}}}
\def\iuv{\ensuremath{I_{\mbox{\tiny UV}}}}
\def\ibulk{\ensuremath{I_{\mbox{\tiny bulk}}}}

\def\iqft{\ensuremath{I_{\mbox{\tiny QFT}}}}
\def\icft{\ensuremath{I_{\mbox{\tiny CFT}}}}
\def\zgrav{\ensuremath{Z_{\mbox{\tiny Grav}}}}
\def\zqft{\ensuremath{Z_{\mbox{\tiny QFT}}}}
\def\grav{\ensuremath{_{\mbox{\tiny Grav}}}}
\def\qft{\ensuremath{_{\mbox{\tiny QFT}}}}
\def\cft{\ensuremath{_{\mbox{\tiny CFT}}}}
\def\bcft{\ensuremath{_{\mbox{\tiny BCFT}}}}
\def\eff{\ensuremath{_{\mbox{\tiny eff}}}}
\def\bulk{\ensuremath{_{\mbox{\tiny bulk}}}}

\def\sc{\ensuremath{^{\mathsf{c}}}}
\def\ib{\ensuremath{I_{\mbox{\tiny edge}}}}

\newcommand{\rir}{\ensuremath{\overline{\p\mathcal{R}}}}
\newcommand{\eir}{\ensuremath{\overline{\p\mathcal{E}}}}
\newcommand{\euv}{
\ensuremath{\mspace{2mu}
\underline{\mspace{-2mu}\p\mathcal{E}\mspace{-2mu}}
\mspace{2mu}}}
\newcommand{\phir}{\ensuremath{\bar{\phi}}}
\newcommand{\phuv}{
\ensuremath{\mspace{2mu}
\underline{\mspace{-2mu}\mathcal{\phi}\mspace{-2mu}}
\mspace{2mu}}}
\newcommand{\Phir}{\ensuremath{\bar{\Phi}}}
\newcommand{\zir}{\ensuremath{\bar{z}}}

\begin{document}

\begin{titlepage}

\vspace*{1cm}
\begin{center} 
\Large \bf Holographic Coarse-Graining: \\
Correlators from the Entanglement Wedge\\
and Other Reduced Geometries
\end{center}

\begin{center}
Alberto G\"uijosa$^{\ast}$,
Yaithd D.~Olivas$^{\ast}$
and Juan F.~Pedraza$^{\dagger}$

\vspace{0.2cm}
$^{\ast}\,$Departamento de F\'{\i}sica de Altas Energ\'{\i}as, Instituto de Ciencias Nucleares, \\
Universidad Nacional Aut\'onoma de M\'exico,
\\ Apartado Postal 70-543, CDMX 04510, M\'exico\\
 \vspace{0.2cm}
$^{\dagger}$
Departament de Física Quàntica i Astrofísica and
  Institut de Ciències del Cosmos\\
Universitat de Barcelona, Martí i Franquès 1, E-08028 Barcelona, Spain
\vspace{0.2cm}

{\tt yaithd.olivas@correo.nucleares.unam.mx, \\
 alberto@nucleares.unam.mx, juanpedraza@icc.ub.edu}
\end{center}

\begin{center}
{\bf Abstract}
\end{center}
\noindent
There is some tension between two well-known ideas in holography. On the one hand, subregion duality asserts that the reduced density matrix associated with a limited  region of the boundary theory is dual to a correspondingly limited region in the bulk, known as the entanglement wedge. On the other hand, correlators that in the boundary theory can be computed solely with that density matrix are calculated in the bulk via the GKPW or BDHM prescriptions, which require input from \emph{beyond} the entanglement wedge. We show that this tension is resolved by recognizing that the reduced state is only fully identified when the entanglement wedge is supplemented with a specific infrared boundary action, associated with an end-of-the-world brane. This action is obtained by coarse-graining through a variant of Wilsonian integration,  a procedure that we  call \emph{holographic rememorization}, which  can also be applied to define other reduced density or transition matrices, as well as more general reduced partition functions. We find an interesting connection with AdS/BCFT, and, in this context, we are led to a simple example of an equivalence between an ensemble of theories and a single theory, as discussed in recent studies of the black hole information problem.

\vspace{0.2in}
\smallskip
\end{titlepage}

\tableofcontents

\section{Introduction and summary}

According to the holographic correspondence \cite{malda,gkp,w}, certain states of  quantum field theories (QFTs) with many degrees of freedom and strong coupling can be alternatively described as semiclassical states of higher-dimensional gravitational theories.   The dynamical `bulk' spacetime is understood to emerge \cite{maldaeternal,swingle,vr,eprer} from the pattern of entanglement of the `boundary' QFT state $\ket{\Psi}$. 

When we only have access to a limited spacelike region $A$ in the the QFT, or equivalently, to its causal diamond (domain of dependence) $D$, the best characterization of our subsystem is given by the reduced density matrix\footnote{This matrix is often denoted $\rho_A$, to emphasize its dependence on the choice of untraced degrees of freedom. In this paper we will refer repeatedly to $\rho$ and its associated causal diamond $D$, entropy $S$, Ryu-Takayanagi surface $\Gamma$ and entanglement wedge $\cE$ (the last three objects will be defined momentarily). To lighten the notation, we  will generally not attach a subindex to these quantities, understanding that, unless otherwise noted, they correspond to some arbitrary but fixed spatial region $A$.}$^{,}$\footnote{For simplicity, throughout the paper we will use the standard terminology and notation that assumes factorizability of the Hilbert space between the QFT degrees of freedom in $A$ and $A\sc$. The more accurate description in terms of Tomita-Takesaki theory is reviewed in \cite{wittenentanglement}.}
\begin{equation}\label{rhoa}
\rho\equiv\tr_{A^\mathsf{c}}(\ket{\Psi}\bra{\Psi})~,
\end{equation}
which will be mixed if the overall state has entanglement between $A$ and its complement $A^\mathsf{c}$. This is quantified by the von Neumann entropy of the reduced state, $S\equiv-\tr(\rho \ln\rho)$,
which is known in this context as the entanglement entropy. In the bulk, it is computed by the celebrated Ryu-Takayanagi (RT) formula \cite{rt,hrt,lm,dlr}, 
\begin{equation}\label{rt}
S=\frac{\mbox{Area}(\Gamma)}{4G_N}~,
\end{equation}
 where $\mbox{Area}(\Gamma)$ denotes the area of
the smallest extremal codimension-two bulk surface $\Gamma$ that can be continuously deformed to $A$, with $\p \Gamma =\p A$. Equation (\ref{rt}) holds when the bulk theory is classical Einstein gravity. Generalizations away from classicality or Einsteinianity can be found respectively in \cite{bdhm2,flm,ew}
and \cite{dong,camps}.

The prominent role played in (\ref{rt}) by the RT surface $\Gamma$ eventually led to the proposal of subregion duality, which asserts that the reduced state corresponds to the bulk region demarcated by this surface. In more detail, the seminal works \cite{densitymatrix,wall,hhlr} conjectured that the reduced density matrix $\rho$ is dual to the entanglement wedge of $A$, denoted $\cE$ and defined as the domain of dependence of any codimension-one bulk spacelike region extending between $A$ and $\Gamma$. A complete $\rho\leftrightarrow\cE$ duality would entail the ability to fully translate in both directions. 

The bulk-to-boundary translation is known as `bulk reconstruction', and seeks to identify the region in the bulk that can be fully reconstructed with the information in $\rho$. Building on important insights gained over the years in \cite{bena,hkll,hmps,bousso1,densitymatrix,wall,pr,morrison,hhlr}, this question was definitively answered in \cite{jlms,dhw,fl,cotler,kim,aal,chen}, which showed that, indeed, all local bulk operators in $\cE$ are fully reconstructible within $A$. Crucial for this achievement was the realization that holography works as a special type of code for quantum error correction (QEC) \cite{adh,harlow,ap}, and bulk reconstruction is meaningful only within a `code subspace' of the QFT, where bulk effective field theory is approximately valid.
Useful reviews on bulk reconstruction can be found in \cite{harlowtasi,dejonckheere,kajuri,bartekreview,ayan}.

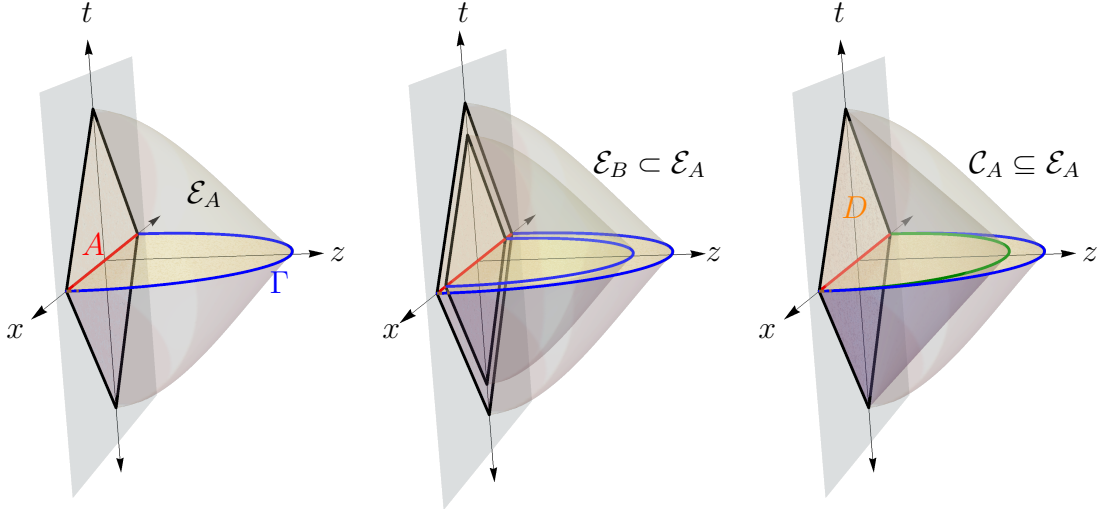
\begin{figure}[!t]
\begin{center}
  \includegraphics[width=5cm]{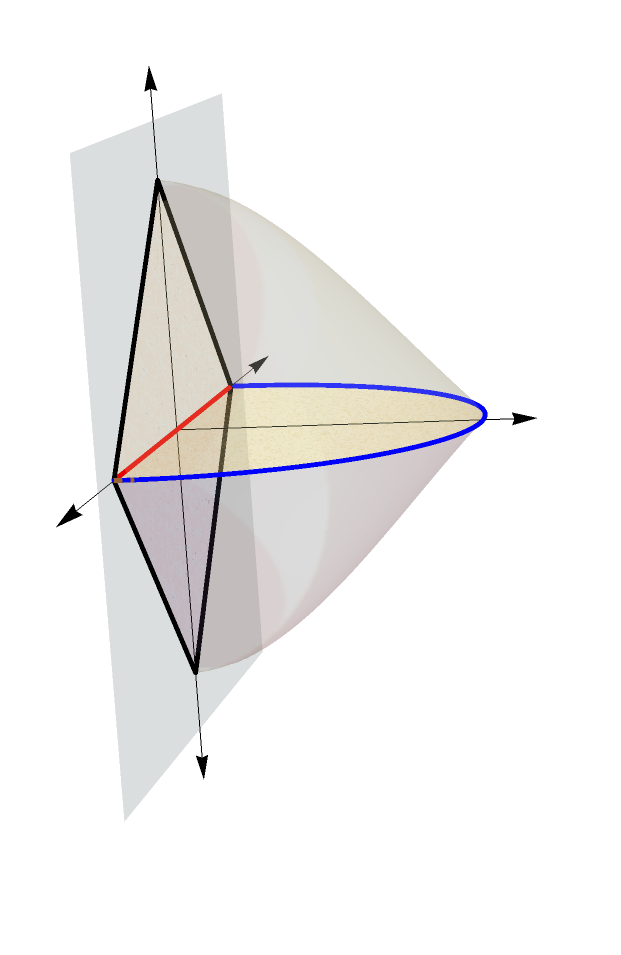}\includegraphics[width=5cm]{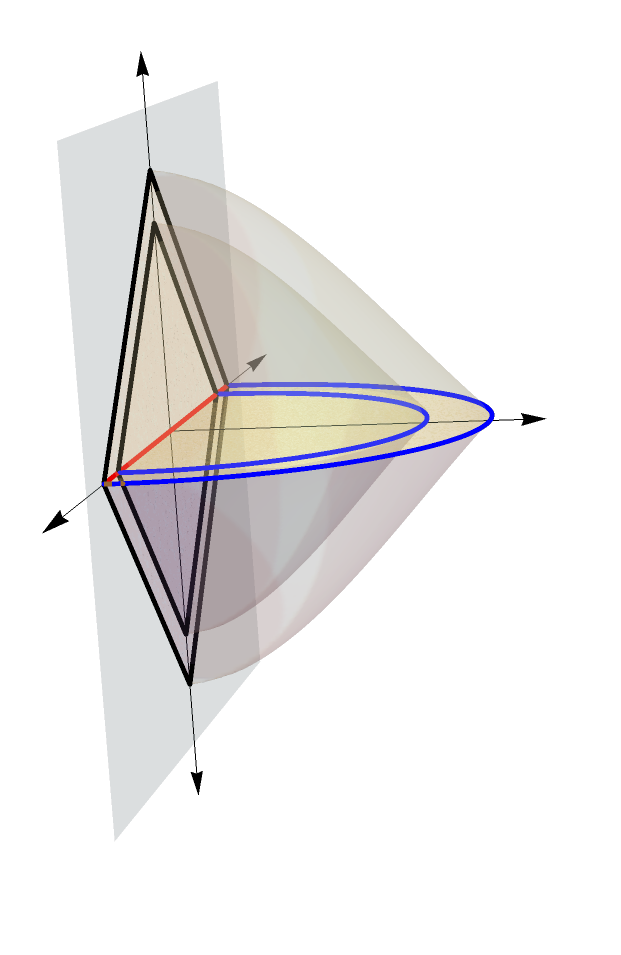}\includegraphics[width=5cm]{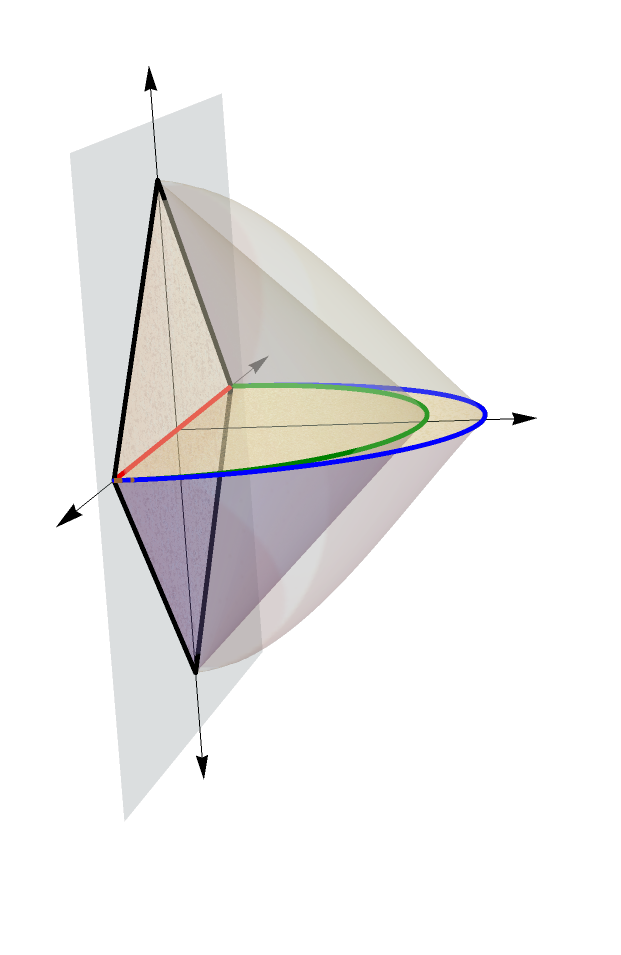}
  \hspace*{-1cm}
  \setlength{\unitlength}{1cm}
\begin{picture}(0,0)
\put(-11.9,5.0){$\cE_{A}$}
\put(-13.3,4.3){\color{red}$A$}
\put(-10.8,3.8){\color{blue}$\Gamma$}
\put(-6.5,5.4){$\cE_{B}\subset\cE_{A}$}
\put(-1.5,5.4){$\cC_{A}\subseteq\cE_{A}$}
\put(-3.2,4.8){\color{orange}$D$}
\put(-14.3,3.2){$x$}
\put(-9.3,3.2){$x$}
\put(-4.3,3.2){$x$}
\put(-13.3,7.4){$t$}
\put(-8.3,7.4){$t$}
\put(-3.3,7.4){$t$}
\put(-10,4.23){$z$}
\put(-5,4.23){$z$}
\put(0,4.23){$z$}
\end{picture}
\end{center}
\vspace*{-0.8cm}
\caption{Left: Schematic portrayal of the entanglement wedge $\cE_{A}$ associated with a given boundary subregion $A$. It is bounded by null geodesics that are shot from the RT surface $\Gamma_A$ towards the boundary. Center: representation of entanglement wedge nesting, a property that implies $B\subset A\Rightarrow \cE_{B}\subset\cE_{A}$. Right: representation of causal wedge inclusion, $\cC_{A}\subseteq\cE_{A}$, where $\cC_{A}$ refers to the bulk region that is causally accessible from $D$, the causal diamond of $A$.
\label{entanglementwedgefig}}
\end{figure}

 Subregion duality implies a number of conditions that the entanglement wedge must satisfy, which follow from properties of $\rho$. This includes entanglement wedge nesting  
(i.e., $B\subset A\Rightarrow \cE_{B}\subset\cE_{A}$) and causal wedge inclusion, $\cC\subseteq\cE$, where $\cC$ refers to the bulk region that is causally accessible from $D$ (when $D$ is viewed as residing on the boundary) \cite{densitymatrix,wall,hhlr}. These properties must hold even beyond classical Einstein gravity, posing several non-trivial constraints that the bulk must satisfy \cite{akll,cmp}. See Fig.~\ref{entanglementwedgefig} for schematic illustrations.

Another crucial property of $\rho$ is the fact that it allows the correct determination of all correlators of local or extended operators $\cO_i$ placed within $A$: 
\begin{equation}\label{correlators}
\tr_A(\rho\,\cO_1\cdots\cO_n)=\bra{\Psi}\cO_1\cdots\cO_n\ket{\Psi}~.
\end{equation}
On the flip side, $\rho$ would not know about correlators of operators that are inserted \emph{outside} of $A$. From the subsystem perspective, the insertion of such operators amounts to a change of the reduced state, and would require recomputation of $\rho$.

In this paper, we study the implications of (\ref{correlators}) for subregion duality. Notice that, contrary to bulk reconstruction, this question goes in the direction of boundary-to-bulk translation: if $\rho$ is indeed dual to $\cE$ to the full extent allowed by the QEC structure, then all instances of (\ref{correlators}) within the code subspace
must be fully encoded in terms of bulk effective field theory on $\cE$. 

In QFTs with a strongly-coupled UV fixed point, large central charge and a sparse spectrum \cite{hpps}, 
correlators of local operators are obtained holographically through the well-known GKPW prescription \cite{gkp,w}, which equates the boundary and bulk partition functions
\begin{equation}\label{zz}
Z_{\mbox{\tiny QFT}}[J_l]
=Z_{\mbox{\tiny Grav}}[\phuv_l=J_l]
\simeq\exp(iI^{\mbox{\tiny on-shell}}_{\mbox{\tiny Grav}}[\phuv_l=J_l])~.
\end{equation}
In this expression, $J_l(x)$ refers to the linear source for $\cO_l(x)$, and $\phuv_l(x)$ is (up to rescaling) the asymptotic boundary value of the bulk field $\phi_l(x,z)$ that is dual to $\cO_l(x)$.  
The last equality in (\ref{zz}) evaluates the partition function of the gravitational theory in a saddle-point approximation, which is usually the only regime where we have quantitative control. The gravity action is holographically renormalized \cite{hsweyl,bkstress,ejm,dhss,skenderis} by counterterms defined at the UV boundary, $I_{\mbox{\tiny Grav}}=I_{\mbox{\tiny bulk}}+I_{\mbox{\tiny UV}}$, and is evaluated on shell in the final step of (\ref{zz}).

The starting point for our work is the observation that (\ref{zz}) seemingly contradicts (\ref{correlators}), because even when the sources $J_l(x)$ are turned off in $A^{\mathsf{c}}$, so that operator insertions are purely within $A$,  $Z_{\mbox{\tiny Grav}}[J_l]$ still requires information from $\cE^{\mathsf{c}}$, the bulk region \emph{beyond} the entanglement wedge. 
This is true even for basic two-point functions of simple operators, well within the confines of the relevant code subspace in the QFT. 
This contradiction reveals that, by itself, the identification of $\cE$ is not equivalent to a complete specification of $\rho$. 

What we need then is a reformulation  of the bulk recipe for correlators in $A$, that makes reference \emph{only} to the entanglement wedge. Given that the issue is tracing out some of the degrees of freedom contributing to the (bulk or boundary) path integral, the natural tool at our disposal is Wilsonian integration, whose holographic implementation was developed in  \cite{hp,flr,bgl}. For our purposes, we will seek to apply it in a nonstandard manner: instead of tracing over a UV region, as is done in standard Wilsonian renormalization, we will trace over the bulk region $\cE^{\mathsf{c}}$, which includes the  IR region as well as what amounts to the UV in $A^{\mathsf{c}}$.

The net result of this approach will be a boundary action $I_{\mbox{\tiny IR}}$ defined at the IR end of the entanglement wedge, whose inclusion guarantees that correlators within $A$ are correctly reproduced. Its presence is analogous to the counterterm action \iuv\ prescribed by holographic renormalization, but whereas the latter serves to cancel the divergences in
\ibulk, the former is needed to keep the memory of the portion of the state encoded by $\cE^{\mathsf{c}}$. For this reason, we refer to our procedure as \emph{holographic rememorization}.  

Conceptually, the issue is that the reduced density matrix $\rho$ specifies not only the subset $A$ of degrees of freedom that remain untraced, but also their state. In the bulk, the identification of $\cE$ goes along with the choice of $A$, but the specification of the state is not complete until the boundary action \iir\ is provided. The full association is therefore not $\rho\leftrightarrow\cE$, but $\rho\leftrightarrow(\cE,\iir)$.

We have thus far focused on the incompleteness, in the context of subregion duality, of the standard GKPW recipe for correlation functions. Alternatively, one can employ the BDHM(BKLT) or `extrapolate' prescription \cite{bdhm,bkextrapolate}, which  extracts boundary correlators from bulk correlators at insertion points that are made to approach the anti-de Sitter (AdS) boundary. This procedure leads to the same results as GKPW \cite{hs}, and is likewise incomplete, because the computation of bulk Green's functions requires specifying boundary conditions.
The correct specification can only be made purely within the entanglement wedge when the boundary action \iir\ is taken into account. 

The need to supplement in this manner the GKPW and BDHM prescriptions, to enable their proper restriction to $\cE$, is connected with the findings of \cite{rindler}. That work identified a challenge to subregion duality similar to the one examined here, by noting that the `hole-ographic' procedure \cite{hole-ography} for reconstructing an arbitrary codimension-2 bulk surface $\gamma$, in terms of `differential entropy' in the boundary QFT, \emph{cannot} be straightforwardly applied within the entanglement wedge, because some of the necessary RT surfaces exit $\cE$. The resolution in \cite{rindler} makes use of the purification that is optimal in the sense of the entanglement of purification \cite{eop}, whose bulk dual was proposed in \cite{takayanagi,phuc}. It was shown in  \cite{rindler} that a slight generalization naturally gives rise to the concept of `differential purification', which, combined with differential entropy, allows full reconstruction of $\gamma$ within $\cE$. Contact with the results in this paper is twofold. On the one hand, the boundary action \iir\ encoding the original $\cE\sc$ would in fact supply the necessary information about the missing RT surfaces, thereby providing a resolution to the puzzle alternative to that of \cite{rindler}. On the other hand, the choice of optimal purification in the QFT can be described in the bulk as a different choice of the IR boundary action for $\cE$, thereby situating the scheme of \cite{rindler} as a particular case of the more general rememorization procedure developed in this paper.  

In completing the statement of subregion duality, our results also relate to  a puzzle raised recently in \cite{bao}. It was noted there that in some situations the bulk metric can be reconstructed far beyond the entanglement wedge, by applying the prescription of \cite{bcfk} to two-dimensional minimal surfaces that reach outside $\cE$ despite being anchored within $A$.  These surfaces can be spanned by string worldsheets that, according to the standard holographic dictionary \cite{reyee,maldawilson}, compute expectation values of Wilson loops that are certainly encoded in $\rho$. The authors of \cite{bao} inferred from this that either there exist data in $\rho$ that determine the metric parametrically far outside $\cE$, conflicting with the standard intuition about subregion duality, or the information of such surfaces is not contained in $\rho$, indicating that there is something wrong with the holographic recipe for Wilson loops. Our perspective is more in line with the first option: by definition, $\rho$ does contain data about the state in $A^{\mathsf{c}}$, and therefore $\cE^{\mathsf{c}}$. However, \emph{it does not fully and uniquely determine that state}. In particular, many states lead to the same reduced density matrix $\rho$ and, hence, the same infrared action \iir.\footnote{Put the other way around, given a reduced state $\rho$, there are infinitely many ways to purify it. Adding to this, one may even consider global states that are mixed to begin with.} Wilson loops whose dual worldsheets exit $\cE$ are directly analogous to the correlators of local operators considered throughout this paper, which are likewise inferred from field profiles that lie partly beyond $\cE$. 
Our results show that the boundary term \iir, which is needed to fully specify $\rho$, 
encodes the external portions of the field profiles, and of the worldsheets relevant to \cite{bao}, thereby 
resolving the tension  with subregion duality. 

We now remark on how our approach differs from previous notions of coarse-graining in holography. The standard Wilsonian RG flow \cite{hp,flr,bgl} of course refers to integrating out the UV region of the QFT, whereas here we apply the same method to carry out a different reduction, largely over IR degrees of freedom. The constructions of \cite{ewcoarse,eflosing}, geared towards explaining area theorems, do aim at leaving out the IR. 
The focus of \cite{ewcoarse} is on reproducing only one-point functions of simple operators,
whereas in our approach, \emph{all} correlators localized within $A$ will be correctly reproduced. In \cite{eflosing}, subregion duality motivates a coarse-graining that requires agreement within a continuous family of finite-size causal diamonds in the boundary theory, in the spirit of \cite{hole-ography}. For the most part, our considerations are confined to one such diamond.\footnote{As explained five paragraphs below, in \S~\ref{extensionsubsec} we do consider situations somewhat similar to those of \cite{eflosing}, but always arriving at a single density matrix, instead of a collection of matrices as in \cite{hole-ography}.} In both \cite{ewcoarse} and \cite{eflosing}, the final bulk region of interest covers the entire spatial extent of the boundary, which is unlike our main interest here.
Also, in both of those papers, the considerations are purely entropic, and therefore purely gravitational, whereas our prescription applies Wilsonian-like integration to all bulk fields, thereby preserving the memory of all correlators. Other related work is \cite{nrs}, which is sort of the opposite of \cite{eflosing}: it focuses on reduction over UV degrees of freedom, again
employing a perspective along the lines of \cite{hole-ography}, that envisions relinquishing access to an infinite collection of infinitesimal regions in position space, 
instead of placing a cutoff in momentum space à la Wilson. Interesting progress along a somewhat similar direction has been made in \cite{barteksewing,bartekquery}.

On the other hand, our approach does have some points of contact with the
AdS/BCFT correspondence \cite{kr,kradsbcft,adsbcft,miao}, the surface-state  correspondence \cite{surfacestate,gwmw}, path integral optimization \cite{pioptimization,takayanagipi}  and tensor networks \cite{swingle,qi,happy,nothappy,walltensor}. 
%%holographic screens
We will comment on them at appropriate points in the text.

This paper is structured as follows. After a brief review of holographic Wilsonian renormalization in \S~\ref{wilsoniansubsec}, we present our rememorization procedure in \S~\ref{rememorizationsubsec}, specifically in Eqs.~(\ref{iraction}) and (\ref{subregion}). The latter is a first version of our main result: an implementation of the GKPW prescription purely within the entanglement wedge $\cE$, consistent with subregion duality. The physical picture is that, after rememorization, the IR boundary  $\eir$ of the entanglement wedge becomes an actual edge of spacetime, where an end-of-the-world (EOW) brane resides, and it is the action $\iir$ of this brane that encodes, in abridged form, the portion of the state that was previously in $\cE\sc$. 

Section~\ref{adsbcftsec} develops the field theory interpretation of rememorization. The analysis applies to any holographic QFT, but for familiarity, we focus on CFTs.  In \S~\ref{spatialsubsec}, we discuss a Wilsonian-like reduction of the CFT in position rather than momentum space, which leaves us with an effective BCFT. This can be used in particular to reduce to a causal diamond $D$, as in (\ref{DiamondInt2}) and Fig.~\ref{reducedcftfig} right. In \S~\ref{translationsubsec}, we relate this to the gravity description via AdS/BCFT, contemplating then a bulk truncated by an EOW brane, as depicted in Fig.~\ref{reducedbulkfig}. 

In \S~\ref{dnsubsec}, we return to (\ref{subregion}) and show that carrying out the $\cD\phir$ path integral over the EOW degrees of freedom has the net effect of imposing a Neumann (or really, nonlocal Robin-like) boundary condition, leading to the second and final form of our prescription, Eq.~(\ref{subregion3}), which should plausibly agree with the bulk dual of the Neumann version of the CFT reduced to $D$, Eq.~(\ref{DiamondInt3}). The equivalence between the bulk setups (\ref{subregion}) and (\ref{subregion3}), or between the boundary setups (\ref{DiamondInt2}) and (\ref{DiamondInt3}), provides a simple example of faithful translation between an ensemble of theories and a single theory, an important property that has been argued for \cite{marolfmaxfield,nomura,neuenfeld}
in connection with the important recent advances in the black hole information paradox \cite{penington,aemm,islands,eastcoast,westcoast,micro}. Interesting constraints on the existence or range of validity of ensembles in a gravitational UV-complete framework have been recently discussed in \cite{mcnamaravafa,heckman}. 

In \S~\ref{extensionsubsec} we discuss, still in general terms, extensions of our prescription. Both in the bulk and in the boundary, one can integrate out to define partition functions reduced to arbitrary spacetime regions. On the other hand, to obtain reduced \emph{states} (expressed through reduced density or transition matrices), one must trace over a well-defined set of degrees of freedom, which entails in particular turning off the corresponding set of sources. If the partial trace is taken in position space, one is naturally led to a reduction to a causal diamond $D$ in the CFT, and its corresponding entanglement wedge $\cE$ in the bulk, with the specific EOW action $\iir$ that arises from rememorizing $\cE\sc$. We discuss a possible generalization of this standard version of subregion duality, obtained by spacetime reduction of the CFT to a region $R\neq D$, and use of AdS/BCFT to identify the dual bulk region $\cR\neq\cE$. One can alternatively trace over momentum modes, with a bulk implementation as portrayed in Fig.~\ref{coarsegrainfig} right. This is somewhat along the lines of \cite{vijaymark}, but is markedly distinct from the hole-ographic approach \cite{hole-ography}, which focuses on a collection of density matrices, instead of a single one.  A separate kind of generalization is to exploit the freedom to employ a different $\iir$, thereby changing the reduced state associated with the leftover region.

The final two sections of the paper are devoted to a couple of concrete applications of rememorization, which explicitly illustrate the basic features. In Section~\ref{scalarsec} we rememorize a free scalar field in pure Poincar\'e-AdS$_{d+1}$, first in \S~\ref{wallsubsec} for a simple example where the  EOW brane is a wall at fixed radial depth, and then in \S~\ref{generalsubsec} for reduction to an arbitrary bulk region. 

In Section~\ref{geodesicsec} we consider the case where the CFT operators of interest have large conformal dimension, allowing the use of the geodesic approximation in the bulk \cite{balasubramanianross,louko}. Geodesics that exit the bulk region of interest provide a good analog of the Wilson loops considered in \cite{bao}, with the advantage of being simple enough to be analytically tractable. We work out the EOW brane action $\iir$  in a couple of examples: a wall at fixed radial depth in \S~\ref{wallgeodesicsubsec}, and the disconnected entanglement wedge for two distant intervals in \S~\ref{doubleintervalsubsec}. Both cases serve to illustrate the equivalence between the Dirichlet and Neumann approaches introduced in \S~\ref{dnsubsec}, with the latter approach proving to be a useful shortcut. 

There are various interesting questions that remain for future work. Explicit rememorization of Wilson loops is an obvious target. In this context, it is worth noting that the resulting EOW brane will necessarily allow strings and other types of extended objects of the given gravitational theory to end on it. 

One area that would benefit from improved understanding is the association between a radial cutoff and a specific set of momentum space sources, in CFTs as well as in more general holographic QFTs \cite{grozdanov,kiritsis,bartekquery}. This association is bound to be even more subtle in the time-dependent case \cite{cesaralbertojuan}.  Another area that deserves to be better understood is the precise translation between the BCFTs discussed here and their dual EOW branes. 

A natural extension is to rememorize the metric, starting with the free graviton field and moving on to the full problem. This is related to the Randall-Sundrum story \cite{randallsundrum,rsverlinde,rsgubser,kr}, which has been much discussed recently in the context of double holography and entanglement islands \cite{islands,gengkarch,myersislands,neuenfeld,ow}, with the concept of island having been called into question for theories with long-range gravity in \cite{gengkarch,karchraju}. Rememorization of the metric involves the familiar complications of gauge invariance, the associated edge modes, and gravitational dressing  \cite{harlowwormhole,donnellyfreidel,jafferis,gomeshopfmullerriello,dongharlowmarolf,akersrath,giddingssplitting,freidel}, as well as the possible perturbative or nonperturbative obstructions to factorizing the metric configuration into two complementary regions \cite{rajunonsplitting,giddingsnonsplitting}, which might for instance impose constraints on the class of spacetime regions for which rememorization is allowed. Knowledge of the metric dependence of the  EOW brane action $\iir$ would in particular allow verification of the expected self-consistency of the brane's backreaction. This boundary action would also serve to codify RT surfaces that exit an entanglement wedge $\cE$, allowing in particular a determination of the specific $\iir$ associated with the purification of $\rho$ that is optimal in the sense of  entanglement of purification \cite{eop,takayanagi,phuc}, mentioned above. This would in turn enable study of that purification and the `excised' version of the duality that directly equates it with the entanglement wedge \cite{rindler}.  
%%In QFT, there is no definite state (ket) for entangled subsystems. In the gravity theory, this relates to lack of complete definition for bulk across RT surface. Need edge modes.

%%Purification: recall interval within BCFT.

\section{Coarse-graining by integrating out geometry}
\label{coarsesec}
\subsection{Brief review of holographic Wilsonian renormalization}\label{wilsoniansubsec}

In a Poincar\'e-invariant $d$-dimensional QFT with spacetime coordinates $x^{\mu}\equiv (t,\vec{x})$ and fields $\Phi(x)$, the standard Wilsonian effective action at floating cutoff $\Lambda$,  $I\qft^{\Lambda}$~, is obtained \cite{wilsonkogut} by integrating out the Fourier modes
$\Phi(p)$ with 
momentum
$p>\Lambda$,
\begin{equation}\label{ilambda}
\exp\left(iI\qft^{\Lambda}[\Phi_{p<\Lambda}]\right)
\equiv\int\cD \Phi_{p>\Lambda}
\exp\left(iI\qft [\Phi]\right)~,
\end{equation}
so that the partition function can be reexpressed as
\begin{equation}\label{zwilson}
    \zqft=\int\cD\Phi \exp\left(i\iqft [\Phi]\right)
=\int\cD \Phi_{p\le\Lambda}
\exp\left(iI\qft^{\Lambda}[\Phi_{p\le\Lambda}]\right)~.
\end{equation}

Let us now briefly review the holographic implementation of Wilsonian integration \cite{hp,flr,bgl}. For simplicity, we consider a bulk scalar field $\phi$ on a fixed $(d+1)$-dimensional asymptotically AdS geometry. In Poincar\'e coordinates $(x,z)\equiv (t,\vec{x},z)$, the pure AdS metric is
\begin{equation}\label{poincare}
ds^2=\frac{L^2}{z^2}\left(-dt^2+d\vec{x}^2+dz^2\right)~,
\end{equation}
so we have in mind a geometry that approaches (\ref{poincare}) as $z\to 0$, possibly with non-normalizable falloff.
Extensions can be made to any locally asymptotically AdS geometry, possibly tensored or warped with an accompanying compact manifold,
as well as to other types of bulk fields (including the metric itself), with or without interactions.
The limit $\gn\to 0$ where backreaction is suppressed corresponds to considering large central charge at the UV fixed point of the QFT, $c\to\infty$.

The well-known UV-IR connection \cite{uvir,pp} relates the bulk radial direction $z$ to an energy scale $1/z$ in the QFT. This is naturally taken to refer to a resolution scale in the sense of the renormalization group (RG) \cite{akhmedov,bkrg,dbvv,grozdanov,kiritsis},
and attempts have been made to derive holography directly from this connection \cite{lee,douglas,shyam,ryu}. For the specific case of the Wilsonian RG, the floating UV cutoff $\Lambda$ is translated into a radial position $z_{\Lambda}\equiv 1/\Lambda$ in the bulk, and the dual description of (\ref{zwilson}) then involves integrating out the UV region of the geometry, $z<z_{\Lambda}$ \cite{hp,flr}.\footnote{More precisely, a bulk radial cutoff $z_{\Lambda}$ corresponds \emph{not} to a sharp cutoff in momentum space, but to a smooth cutoff akin to \cite{polchinski}. We will have more to say about this point in \S~\ref{extensionsubsec}. \label{nonsharp}}  
In more detail, the path integral that computes the partition function on the gravity side is separated in the following form:
\begin{align} \label{uvintegration}
 Z\grav[\phuv] 
 &=\int \mathcal{D}\phi  \exp\left(i(\ibulk[\phi]+\iuv^{\eps}[\phuv])\right) \nonumber\\
&=\int \cD\phi_{z<z_{\Lambda}}
     \cD\phir
     \cD\phi_{z>z_{\Lambda}}
    \exp\left(i(\ibulk[\phi_{z<z_{\Lambda}}]
    +\iuv^{\eps}[\phuv])\right) \exp\left(i\ibulk[\phi_{z>z_{\Lambda}}]\right)
            \nonumber\\
&\equiv\int 
     \cD\phir
     \cD\phi_{z>z_{\Lambda}}
     \exp\left(i(\ibulk[\phi_{z>z_{\Lambda}}]
     +\iuv^{z_{\Lambda}}[\phir,\phuv])\right)~.
\end{align} 
Since we are using the Lorentzian version of the correspondence \cite{bklorentzian,svr}, the path integral includes a specification of the initial and final states, which are left implicit for now. In the first line of (\ref{uvintegration}), we have taken into account the usual UV counterterms $\iuv^{\eps}$ needed for holographic renormalization \cite{hsweyl,bkstress,ejm,dhss,skenderis}, which are defined at the original cutoff surface $z=\epsilon$.
It is implicitly understood that for the time being we are working with the standard boundary condition
$\phi(x,z)\to z^{d-\Delta}\phuv(x)$ as $z\to\eps$,
where $\phuv(x)$ is to be equated with the QFT source $J(x)$ that couples linearly to the local operator $\cO(x)$ of interest, and $\Delta$ is the scaling dimension of $\cO$.
In the second line of (\ref{uvintegration}), $\phir$ denotes the value of the field at the floating cutoff surface $z=z_{\Lambda}$. In the final line, the new boundary term $\iuv^{z_{\Lambda}}$ has been generated by the UV integration.

\subsection{Holographic rememorization}
\label{rememorizationsubsec}

Contemplating the split path integral in the second line of (\ref{uvintegration}),  it is evident that we can exchange the roles of the UV and IR, so that we instead choose to integrate out the IR region, $z>z_{\Lambda}$.
Upon doing so, we arrive at 
\begin{equation} \label{irintegration}
 Z\grav[\phuv] 
=\int\cD\phir
     \cD\phi_{z<z_{\Lambda}}
     \exp\left(i(\ibulk[\phi_{z<z_{\Lambda}}]
     +\iir[\phir]+\iuv[\phuv])\right)~,
\end{equation} 
where we now have an \emph{infrared} boundary term \iir. Evidently, the QFT interpretation is that we are now defining an upward Wilsonian RG flow, 
integrating out the field modes with $p<\Lambda$.\footnote{See the previous footnote.} 
In (\ref{irintegration}) we have no longer labeled $\iuv$ with the location $z=\epsilon$ of the UV cutoff surface, which is now held fixed. For the same reason,  in the remainder of the paper we will omit the $\eps$ label in all mention of the UV counterterms.

In both (\ref{uvintegration}) and (\ref{irintegration}), we are carrying out the path integral up to a constant radial depth $z=z_{\Lambda}$. This cutoff surface can be turned into a more general timelike surface via an $x$-dependent reparametrization of $z$, which is known to correspond to a Weyl transformation in the QFT \cite{isty}. So if we carry out the path integral up to an $x$-dependent depth, $z=\zir(x)$, we will be considering a spacetime-dependent RG flow, as in \cite{osborn,melosantos}.

\subsubsection{Reducing to the entanglement wedge}
\label{wedgesubsubsec}
The preceding observation suggests a natural way to address the challenge described in the Introduction. To obtain a generalization of the GKPW recipe that is defined purely within the entanglement wedge $\cE$ associated with a spatial region $A$ in the QFT, we ought to integrate out the entire complementary region $\cE\sc$. 

To spell this out, let $D$ refer to the causal diamond of $A$ in the QFT, and denote 
 the profile of the source in $D$ as $J_D(x)$, with   $J_{D\sc}(x)$ standing then for the source in the complement $D\sc$. All correlators within $D$ in a given global state $\ket{\Psi}$ are encoded in $\zqft[J_D,\,J_{D\sc}\!\!=\!\!0]$, so in the bulk we are interested in 
$\zgrav[\phuv_D,\,\phuv_{D\sc}\!\!=\!\!0]$.\footnote{Nothing stops us from turning on the sources $J_{D\sc}$, but that calculation would amount to having an adjustable reduced state, and the infrared action (\ref{iraction}) would naturally be a functional of $J_{D\sc}$. We would then be able to compute \emph{all} correlators, independently of whether the operators are inserted in $D$ or $D^{\mathsf{c}}$. That is \emph{not} the case when we only have access to a specific reduced density matrix $\rho$ describing the state on $D$, which is situation of main interest in this paper. We will return to this point in \S~\ref{extensionsubsec}.} 
These boundary conditions are implicitly understood to hold in the expressions below.
We will define
\begin{equation} \label{iraction}
\exp\left(i(\iir[\phir])\right)
\equiv\int \cD\phi_{\cE\sc}
     \exp\left(i(\ibulk[\phi_{\cE\sc}])\right)~,
\end{equation}
where $\phir(x)\equiv\phi(x,\zir(x))$ 
is the value of the bulk field on $\eir$,
the interface between $\cE$ and $\cE\sc$. Notice that the counterterm action $\iuv[\phuv_{D\sc}]$ drops out, since we have turned off the source in that region, and the subleading behaviour $\phi\,\propto\, z^{\Delta}$ is the normalizable mode that leads to a finite bulk action. The source within $D$, $\phuv_D=J_D$, plays no role in the right-hand side of (\ref{iraction}), so by construction $\iir$ is independent of it. 

Our prescription for computing correlators within subregion duality is to employ the partition function
\begin{equation} \label{subregion}
 Z\grav[\phuv] 
=\int\cD\phir
     \cD\phi_{\cE}
     \exp\left(i(\ibulk[\phi_{\cE}]
     +\iir[\phir]+\iuv[\phuv])\right)~,
\end{equation} 
where it is understood from now on that $\phuv$ refers solely to a source within $D$. 
Eq.~(\ref{subregion}) is thus the generating functional for correlators within $D$. The presence here of the IR boundary action is absolutely crucial to comply with (\ref{correlators}), which is a defining property of the reduced density matrix. We thus see that \emph{the information encoded in $\rho$ is dual not merely to the entanglement wedge, but to $\cE$ equipped with the specific IR boundary term $\iir$}.\footnote{In the partition function (\ref{subregion}) or  the correlators it encodes, one has the reduced density matrix inside a trace, as in (\ref{correlators}). To obtain $\rho$ directly, one must cut open the path integral across the time slice $t=t_0$ on which $A$ resides. Strictly speaking, this will only yield a density matrix if the configuration is symmetric under time reversal about $t_0$. We will say more about this point in 
\S~\ref{extensionsubsec}. \label{cutopenfootnote}}  
Physically, the point is that, after tracing over $\cE\sc$, the IR boundary of the entanglement wedge, $\eir$,
becomes an actual edge of spacetime, with very specific dynamics for the end-of-the-world (EOW) brane that resides there. This point will become clearer towards the end of the following section. 

The preceding construction has been formulated for simplicity in terms of a single bulk scalar field $\phi(x,z)$, which is what we need if we restrict attention solely to correlators of its dual scalar operator $\cO(x)$. The same reduction can be performed singly or jointly on other bulk fields, including the metric itself (dealing appropriately with the associated diffeomorphism invariance, see, e.g., \cite{harlowwormhole,donnellyfreidel,jafferis,gomeshopfmullerriello,dongharlowmarolf,akersrath,giddingssplitting,freidel}). 
A combined analysis of our scalar field and the metric is necessary if we are not working in the limit of strictly infinite central charge, and need then to consider how $\phi$ backreacts on the geometry. 

\section{Field theory interpretation,
and connection with AdS/BCFT} \label{adsbcftsec}

We would now like to understand the field-theoretic interpretation of the bulk rememorization procedure developed in the previous section. For this purpose, it will be convenient to begin in Section \ref{spatialsubsec} with a discussion purely within the boundary theory, and return to the bulk later, in \S~\ref{translationsubsec}. For familiarity with the resulting initialisms,  we will phrase our discussion in terms of a conformal field theory (CFT), even though what we will say in \S~\ref{spatialsubsec} applies to any QFT, and \S~\ref{translationsubsec}  \emph{et seq.} would be expected to hold for any holographic QFT. 

\subsection{Spatial coarse-graining in the boundary} \label{spatialsubsec} 

As briefly reviewed in  section \ref{wilsoniansubsec}, the Wilsonian effective action in a CFT$_{d}$, $\icft^{\Lambda}$, is obtained by integrating over Fourier modes above a floating UV cutoff, i.e., $\Phi(p)$ with $p>\Lambda$, where $\Phi$ refers collectively to all fundamental fields in the theory. Of course, we are not confined to integrating only over momentum degrees of freedom. More generally, we can divide the variables of the path integral into any two complementary subsets 1 and 2, and integrate over one of the two sets. This includes the case where we split the QFT into complementary \emph{spacetime} regions $R$ and $R^\mathsf{c}$, and integrate over $\Phi(x)$ with $x\in R^\mathsf{c}$.

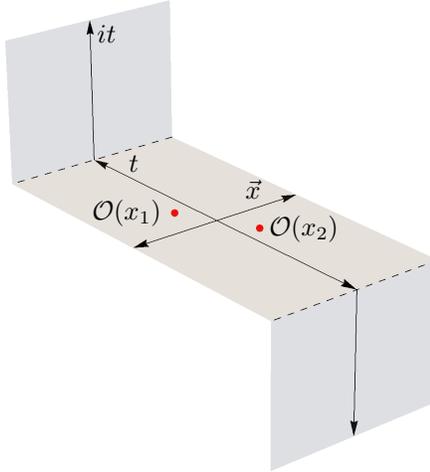
\begin{figure}[t!]
    \centering
    \hspace{-0.2cm}
    \includegraphics[width=7cm,trim={0 2cm 2cm 0},clip]{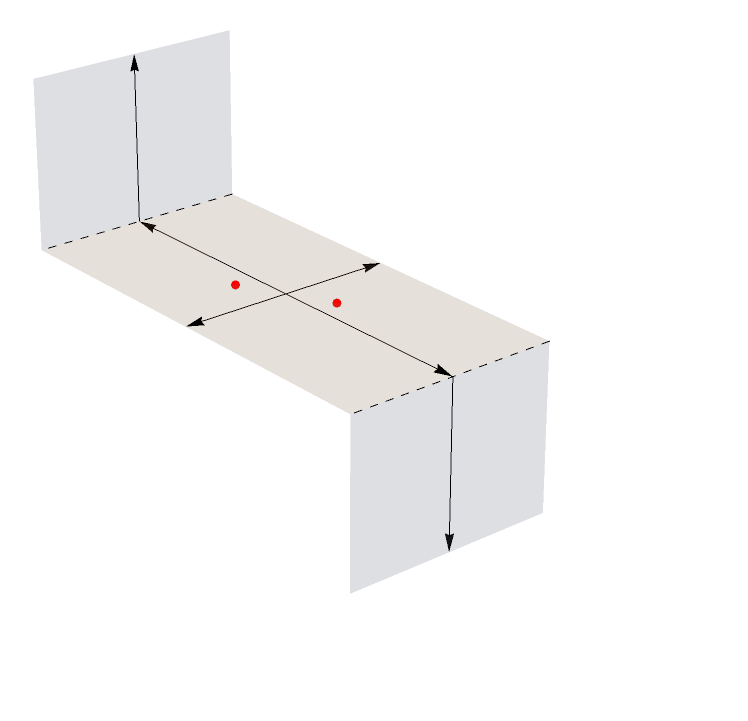}
     \vspace{-0.3cm}
     \begin{picture}(0,0)
\put(-102,106){\footnotesize $\vec{x}$}
\put(-146,116){\footnotesize $t$}
\put(-158,165){\footnotesize $it$}
\put(-160,98){\footnotesize $\mathcal{O}(x_1)$}
\put(-93,92){\footnotesize $\mathcal{O}(x_2)$}
\end{picture}
    \caption{In the figure, we represent the spacetime over which the CFT path integral is carried out, including the complex contour of integration in time.  The Euclidean segments are represented by vertical planes, while the Lorentzian segment corresponds to the horizontal plane. }
    \label{fig:timecountour}
\end{figure}

For concreteness, consider this type of spacetime reduction in the context of  
a real-time $n$-point function in the vacuum $\ket{\Omega}$ of the CFT, 
$\bra{\Omega}{\mathcal{T}}\cO(x_{1})\dotsc \cO(x_{n})\ket{\Omega}$.
 Its standard path-integral computation involves the introduction of a complex time contour, with
one Lorentzian and two Euclidean segments, as in Fig.~\ref{fig:timecountour}. The Euclidean regions prepare the initial and final states.\footnote{Equivalently, we can focus only on the Lorentzian path integral 
$$
\label{LorenztianInt}
\bra{\Phi_{+},T}\cO(x_{1})\ldots \cO(x_{n})\ket{\Phi_{-},-T}
=\int_{\Phi(-T,\vec{x})=\Phi_{-}(\vec{x})}^{\Phi(T,\vec{x})=\Phi_{+}(\vec{x})}
\mathcal{D}\Phi(x)\, \cO(x_{1})\ldots\cO(x_{n})e^{i\icft[\Phi]}~,
$$
with $T>|x^0_j|$ for all $j$, and convolve with the vacuum wavefunctionals $\braket{\Phi_{-},-T}{\Omega}$ and $\braket{\Omega}{\Phi_{+},T}$ to obtain the desired vacuum correlator. Either way, one can of course consider more general matrix elements among states that are not necessarily the vacuum.}

\begin{figure}[t!]
\begin{center}
  \hspace{0.4cm}\includegraphics[width=7cm,trim={0 2cm 2cm 0},clip]{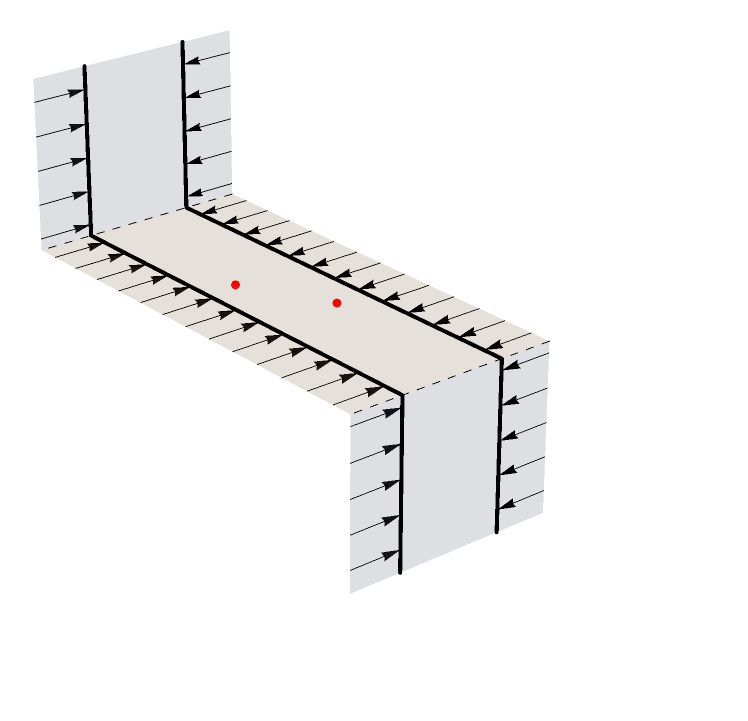} \hspace*{0.5cm}\includegraphics[width=7cm,trim={0 2cm 2cm 0},clip]{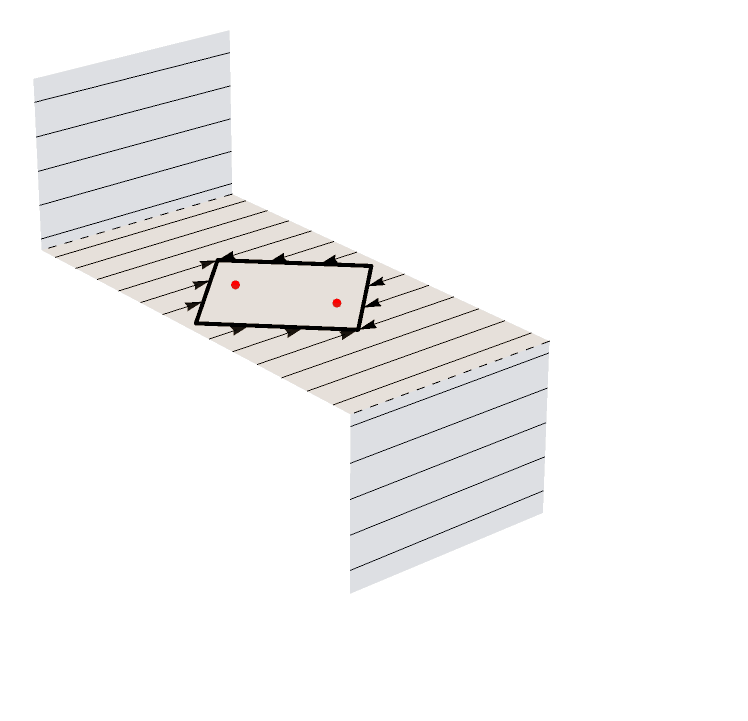}
\end{center}
\begin{picture}(0,0)
\put(-109,138){\footnotesize $x$}
\put(-150,147){\footnotesize $t$}
\put(-164,194){\footnotesize $it$}
\end{picture}
\vspace*{-0.3cm}
\caption{Left: Wilsonian reduction over the complement of a spacetime strip $R=\{x^{\mu}:0\le x^1\le \ell\}$. Right: Wilsonian reduction over the complement of the causal diamond $R=D$ associated with a spatial region $A$.}
\label{reducedcftfig}
\end{figure}
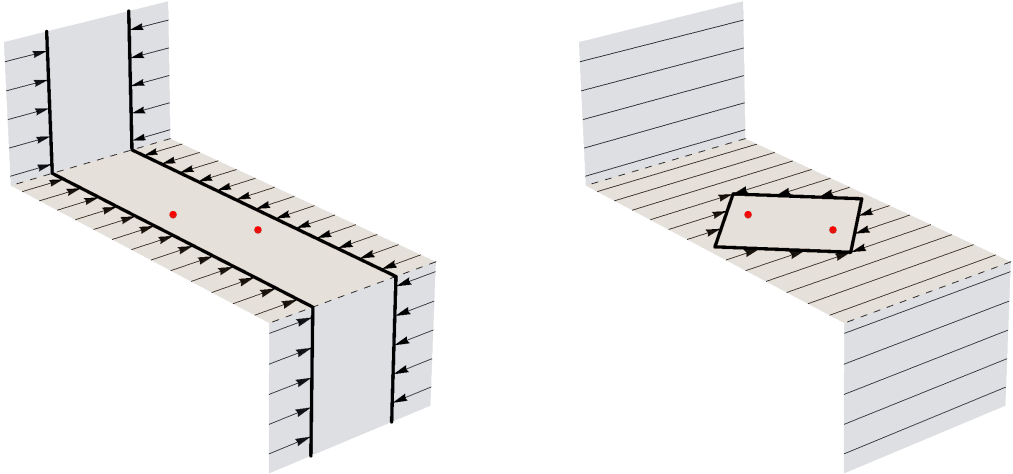

Starting from the full path integral in Fig.~\ref{fig:timecountour}, we can, for instance, reduce to the spacetime strip $R=\{x^\mu:0\le x^1\le \ell\}$, as shown in Fig.~\ref{reducedcftfig} left. This yields
\begin{align}
\label{StripInt}
\bra{\Omega}\mathcal{T}\cO(x_{1})\ldots 
\ket{\Omega}
=\int\mathcal{D}
\Phir_{0}\mathcal{D}\Phir_{\ell}
\mathcal{D}\Phi_{(0,\ell)} \, \cO(x_{1})\ldots
e^{i\icft[\Phi_{(0,\ell)}]
+i\ib[\Phir_{0}]+i\ib[\Phir_{\ell}]}~,
\end{align}
where the subindex of the integration variables denotes the value or range of $x^1$ involved, $\ib$ is the contribution to the effective action arising from integration over $R^\mathsf{c}$, and we have assumed that all operators are inserted within $R$. 

The point to notice in (\ref{StripInt}) is that, within the $\Phir_{0}$  and $\Phir_{\ell}$ path integrals, we are left with a CFT on $R$, a manifold with boundary, with a specific boundary action $\ib$, and Dirichlet boundary conditions $\Phi|_{x^1=0}=\Phir_{0}$, $\Phi|_{x^1=\ell}=\Phir_{\ell}$.
The original correlator is thus recast in terms of a superposition, or ensemble, of boundary conformal field theories (BCFTs).\footnote{For simplicity, we will call these theories BCFTs even when the enforced boundary conditions break all of the  conformal symmetry.} Given a BCFT defined on a $d$-dimensional manifold $ R$, we will refer to $\p R$ as the `edge' and $R\setminus\p R$ as the `interior' of $R$, to avoid confusion with the `bulk' and `boundary' terminology employed customarily in AdS/CFT.

\begin{figure}[!t]
\begin{center}
  \includegraphics[width=7cm]{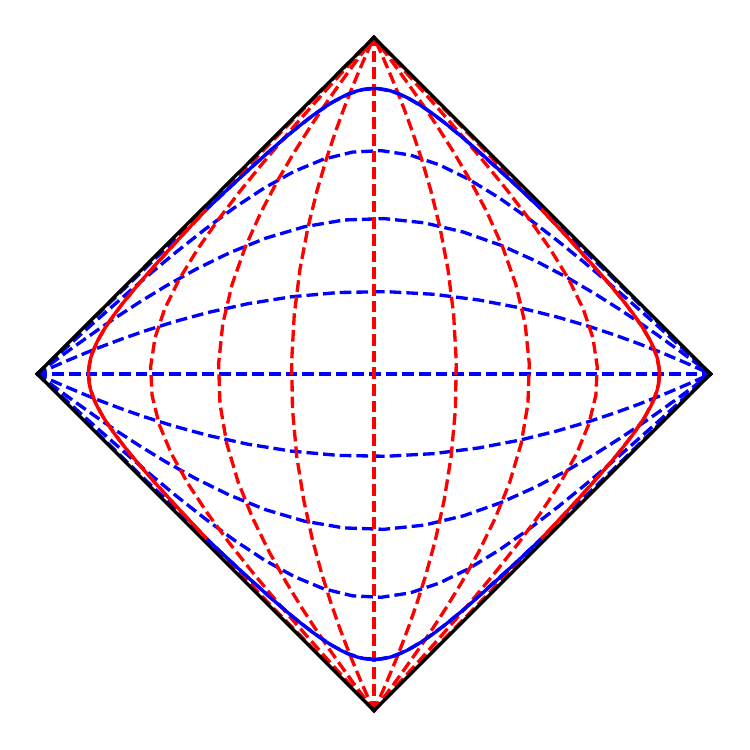} 
\end{center}
\vspace*{-1cm}
\caption{Regularization of a causal diamond $D$, which includes two timelike boundaries, or stretched horizons (depicted in solid red), and two spacelike portions where the initial and final states are specified (depicted in solid blue). \label{fig:rindler}}
\end{figure}

 Clearly, an analogous rewriting can be carried out for other choices of $R$. Of particular interest to us is the case where we reduce to the causal diamond $D$ of some spatial region. See Fig.~\ref{reducedcftfig} right. Integration over $D\sc$ will again produce a boundary term that encodes the information of the integrated region,
 \begin{align}
\label{DiamondInt}
\bra{\Omega}\mathcal{T}\cO(x_{1})\ldots\ket{\Omega}
=\int\mathcal{D}\Phir
\mathcal{D}\Phi_{D} \, \cO(x_{1})\ldots e^{i\icft[\Phi_{D}]
+i\ib[\Phir]}~.
\end{align}
 Here again, within the $\Phir$ integral, we find a BCFT with a Dirichlet boundary condition on $\p D$. This setup is evidently time-dependent: our field theory lives on a region that first expands and then contracts at the speed of light. This can be understood as a limit of a region $R$ where we smooth out the edges of $D$, as in Fig.~\ref{fig:rindler}. The expansion and contraction then proceeds subluminally, and there are two spacelike portions of $\p R$, where the boundary conditions specify the initial and final states. This perspective dispels any worries that the Big Bang-like and Big Crunch-like singularities at the temporal tips of $D$ might make our null BCFT ill-defined.  For use below, we note that all correlators of the form (\ref{DiamondInt}) can of course be repackaged in terms of a generating functional,
 \begin{align}
\label{DiamondInt2}
Z_{\bcft}[J]=
\int\mathcal{D}\Phir
\mathcal{D}\Phi_{D} \, 
\exp\left(i(\icft[\Phi_{D}]
+\ib[\Phir]+\int_{D}\cO(x)J(x))\right)~.
\end{align}
 
\subsection{Translation to the bulk} \label{translationsubsec}

Having made contact with a BCFT through the spacetime
Wilsonian reduction of the preceding subsection, and assuming the existence of a holographic description, we can invoke the standard AdS/BCFT duality \cite{kr,kradsbcft,adsbcft,miao}. 

The generic expectation is for such a BCFT to correspond holographically to a $(d+1)$-dimensional bulk (possibly tensored or warped with an accompanying compact manifold) with the same asymptotics as the geometry dual to the CFT in the interior, capped off by an end-of-the-world (EOW) brane anchored on the edge. In other words, the edge of spacetime in the boundary is extended into the bulk. For simplicity, the EOW brane is regarded as a strictly $d$-dimensional object where spacetime terminates abruptly \cite{kr,kradsbcft,adsbcft,miao}, but this might be just an effective description of a higher-dimensional and possibly more gradual degeneration in a stringy setting, as in \cite{dhoker,aharony,bachas,dhoker2}. Just like the existence of a geometric description of the bulk is made possible by a particular class of patterns of entanglement \cite{hhm,entropycone,hernandezcuenca,veronikamukundcone} in the CFT interior, the existence of the EOW brane as a well-defined object is expected to arise from a restricted type of entanglement involving the edge degrees of freedom.\footnote{Very recently, this question has been examined in \cite{adsbcftnotgeneric} from the perspective of looking for (approximate) singularities in the BCFT correlators that signal the presence of the EOW brane. These singularities were argued to be very non-generic among BCFTs. By construction, the BCFTs under consideration in this paper reproduce the correlators in the full CFT prior to the Wilsonian reduction, and therefore do not possess such singularities. In the bulk, the boundary action $\iir^{\mbox{\tiny N}}$ that we will settle on at the end of the next subsection ensures that, in the terminology of \cite{adsbcftnotgeneric}, our EOW branes have infinite causal depth.}

The precise nature of the bulk, and the action and location of the EOW brane, naturally depend on the specifics of the  BCFT. The original AdS/BCFT papers considered the simplest proof-of-concept case where the brane action is just a constant-tension term, and the bulk is still pure AdS, with the backreaction of the brane  determining the precise depth at which the bulk is cut off. More complicated actions and backreactions would be needed to encode BCFTs with various edge conditions and edge actions.

\begin{figure}[!t]
\begin{center}
  \includegraphics[width=7.5cm]{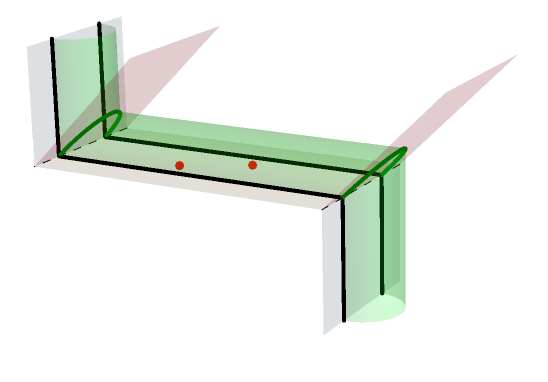}\includegraphics[width=7.5cm]{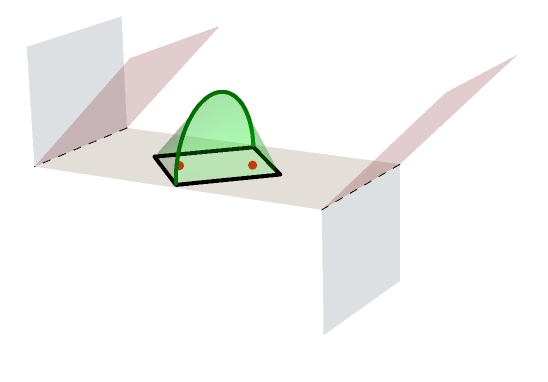}
\end{center}
\vspace*{-1.3cm}
\caption{Holographic dual of the Wilsonian reduction over the complement of: a spacetime strip (left figure, corresponding to Fig.~\ref{reducedcftfig} left) and a causal diamond (right figure, corresponding to Fig.~\ref{reducedcftfig} right). In both figures, the end-of-the-world brane is depicted in green, and the red planes represent surfaces along which the Euclidean and Lorentzian bulk portions are glued together (with the juncture conditions specified in \cite{svr}). \label{reducedbulkfig}}
\end{figure}

In our setup, the original, full CFT calculation is known to be dual to a gravitational computation in a geometry with one Lorentzian and two Euclidean regions, with appropriate matching conditions at the junctures \cite{svr}. In the field theory, we then reformulate the computation in terms of an ensemble of BCFTs on a specific manifold $ R$, with the physics of the edge $\p R$ entirely determined by the spacetime Wilsonian integration described in the previous subsection. For the strip and the causal diamond, the dual configurations are depicted schematically in Figs.~\ref{reducedbulkfig} left and right. More precisely, each bulk figure instantiates the dual of a particular BCFT, and the remaining integral over Dirichlet edge conditions $\Phir$ gives rise then to a superposition of different EOW branes, living on different bulk geometries.\footnote{Here, we use the words  `geometry' and `brane' in a loose sense: for an arbitrary choice of $\Phir$, the BCFT state defined by the $\cD\Phi_R$ path integral might not have the appropriate entanglement structure  \cite{hhm,entropycone,hernandezcuenca,veronikamukundcone} to be encoded in a smooth metric capped off by a localized brane.}

This bulk configuration bears some resemblance with the one resulting from the holographic `rememorization' procedure defined in Section \ref{rememorizationsubsec}. Both involve Wilsonian integration and lead to EOW branes. In both cases, the edge conditions remain to be summed over. But there are also some differences. These partly arise from the simplified discourse we have adopted in both descriptions. E.g., a single, decoupled bulk field would encode correlators only of the dual gauge-invariant operator, whereas a CFT path integral with Dirichlet edge conditions for the fundamental fields would allow computation of all correlators, and would not in itself be invariant with respect to the gauge transformations of the full theory
\cite{harlowwormhole,donnellyfreidel,jafferis,gomeshopfmullerriello}.
A more significant consideration is that \emph{a priori} there is no general reason why an arbitrary spacetime region in the CFT will necessarily be dual to a well-defined spacetime region in the bulk. 
We will return to this point in the following two subsections. 

\subsection{Dirichlet vs.~Neumann: ensemble vs. single theory} \label{dnsubsec}

Both in the bulk and in the boundary, we have integrated out degrees of freedom leaving behind in (\ref{subregion}) and (\ref{DiamondInt}) an EOW or edge integral, $\int\cD\phir$ or $\int\cD\Phir$, that still remains to be carried out. Let us now examine what happens when we do so.

For definiteness, we will refer to the situation in the bulk setup, Eq.~(\ref{subregion}), which we repeat here for ease of consultation,
\begin{equation} \label{subregion2}
 Z\grav[\phuv] 
=\int\cD\phir
     \cD\phi_{\cE}
     \exp\left(i(\ibulk[\phi_{\cE}]
     +\iir[\phir]+\iuv[\phuv])\right)~.
\end{equation}
We expect this partition function to relate specifically to the case where the CFT spacetime is reduced to the corresponding causal diamond $D$.

Consider first the limit of infinite central charge in the CFT,  implying  $\gn\to 0$ in the bulk, where the action 
becomes quadratic, 
\begin{equation}
\label{kleingordon}
     \ibulk[\phi]=-\frac{1}{2}
     \int dz d^{d} x \sqrt{-g}
     \left[g^{mn}\partial_{m}\phi\partial_{n}\phi
     +M^2 \phi^2 \right]~,
\end{equation}
and the saddle-point approximation becomes exact. Carrying out the Gaussian integral in (\ref{iraction}), we obtain a quadratic infrared boundary action, 
\begin{equation} \label{iraction2}
 \iir[\phir]=
 \frac{1}{2}
 \int_{\eir
 }\! d^d x\sqrt{-h(x)}
\int_{\eir
}\!  d^d y\sqrt{-h(y)}
 \,\mathcal{K}(x,y)\phir(x)\phir(y)~,
\end{equation} 
with 
$\eir$
the interface between $\cE$
and $\cE\sc$,
$h$ the induced metric\footnote{We have in mind here a regularized (stretched horizon) version of the null IR boundary of $\cE$. This is analogous to the standard UV regularization at $z=\eps$. 
The case of a strictly null boundary  must be addressed as in \cite{null}.\label{stretchedhorizon}}
on $\eir$,
and
$\mathcal{K}$ some bilocal kernel (the explicit expression will be worked out in Section \ref{scalarsec}). The nonlocal character of this EOW action is a general outcome of rememorization, which is of course due to the integration process, and is featured as well, for the same reason, in the edge action of the BCFT. 

Observe now that the variational principle for $\ibulk+\iir$ yields a boundary term
\begin{equation} \label{boundaryterm}
 \int_{\eir
 }\! d^d x\sqrt{-h(x)}
\left(n^m\p_m\phi(x)
-\int_{\eir
}\!  d^d y\sqrt{-h(y)}
 \,\mathcal{K}(x,y)\phi(y)\right)\delta\phi(x)~,
\end{equation} 
with $n$ the outward-pointing unit\footnote{See the previous footnote.} normal.  
By itself, (\ref{boundaryterm}) implies that there are only two choices for the boundary condition that can be consistently imposed on our field:
\begin{eqnarray}
{}&{}&\mbox{Dirichlet (IR)}\qquad\qquad\;\;\;
\phi(x,z)|_{\eir
}=\phir(x)~,
\label{dirichlet}
\\
{}&{}&\mbox{Neumann (IR)}\qquad
n^m\p_m\phi(x,z)|_{\eir
}
=\int_{\eir
}\!  d^d y\sqrt{-h(y)}
 \,\mathcal{K}(x,y)\phir(y)~.
\label{neumann}
\end{eqnarray}
For familiarity, we refer to the second choice as Neumann, even though it is really a nonlocal generalization of a Robin boundary condition.

Within the integrand of $\int\cD\phir\,$, we are of course enforcing the Dirichlet boundary condition (\ref{dirichlet}), with $\phir(x)$ an arbitrary function that we are meant to sum over. When we carry out the Gaussian path integral over $\phir(x)$, we extremize $\ibulk+\iir$ with respect to 
$\delta\phi(x,z)=\delta(z-\bar{z}(x))\delta\phir(x)$,
which leads us to enforcing the Neumann boundary condition (\ref{neumann}). The function $\phir(x)$ is then determined by the normal derivative at 
$\eir$, which will of course vary within the remaining path integral, $\int\cD\phi_{\cE}$. With this understanding, the latter path integral is still weighted by the exponential seen in (\ref{subregion2}), with $\iir$ the \emph{same}  infrared boundary action. The passage here from Dirichlet to Neumann boundary conditions is analogous to the one discussed in \cite{comperemarolf} for the metric at the UV boundary (a setup where further progress has been made recently in \cite{mateos}). 

Moving on to the case of large but finite central charge in the CFT, implying small \gn\ in the bulk, the path integral over $\phi(x,z)$ can be as usual treated perturbatively.\footnote{And, concurrently, we need to start taking into account the backreaction of $\phi$ on the metric.}
Upon integrating out $\cE^{\mathsf{c}}$, we are left with an interacting EOW brane action $\iir$, complementing the interacting $\ibulk$ in $\cE$. At this point, we are meant to enforce the Dirichlet boundary condition (\ref{dirichlet}).
Carrying out $\int\cD\phir$ is no longer equivalent to saddle-point evaluation, but the result can be defined to be the exponential of a new infrared boundary action, $\iir^{\mbox{\tiny N}}$. 
The IR boundary condition for the remaining integral, $\int\cD\phi_{\cE}$, must be compatible with the variational principle associated with $\ibulk+\iir^{\mbox{\tiny N}}$ (see, e.g., \cite{brinkhenneaux}), and the correct choice will be of Neumann type, similar to (\ref{neumann}).  

Overall, we are left then with the \emph{purely UV} path integral
\begin{equation} \label{subregion3}
 Z\grav[\phuv] 
=\int \cD\phi_{\cE}
     \exp\left(i(\ibulk[\phi_{\cE}]
     +\iir^{\mbox{\tiny N}}[\phir]+\iuv[\phuv])\right)~,
\end{equation} 
where both boundary terms perform double duty. The standard counterterm action $\iuv$ eliminates the UV divergences while maintaining compatibility with the usual Dirichlet boundary condition that introduces the CFT source $\phuv$,
\begin{equation}
\mbox{Dirichlet (UV)}\qquad\qquad\;\;\;
\phi(x,z)|_{\euv}=\eps^{d-\Delta}\phuv(x)~,
\label{dirichletuv}
\end{equation}
where the UV cutoff surface $\euv$ is understood to be at $z=\eps$. The rememorization action $\iir^{\mbox{\tiny N}}$ weighs the path integral appropriately to keep the memory of the bulk region $\cE\sc$ that has been entirely eliminated by integrating it out, and enforces the correct, Neumann (or really, Robin-type) boundary condition,
\begin{equation}
 \mbox{Neumann (IR)}\qquad\qquad
n^m\p_m\phi(x,z)|_{\eir}
=\frac{\delta\iir^{\mbox{\tiny N}}}{\delta{\phir(x)}}~,
\label{neumannir}   
\end{equation}
with $\eir
$ the location of the EOW brane, $z=\zir(x)$. Aside from ease of computation, the main difference between the quadratic and non-quadratic cases is just whether $\iir^{\mbox{\tiny N}}$ does or does not coincide with the original Dirichlet action $\iir^{\mbox{\tiny D}}\equiv\iir$.

There are two features here that are worth emphasizing. The first is that performing the integral over $\phir$ in the gravity theory or $\Phir$ in the field theory, aside from transmuting our original Dirichlet boundary/edge condition into a Neumann one, moves us from an ensemble of theories to a single theory. The equivalence between these two perspectives presents us then with a simple example of a crucial property that has been argued for \cite{marolfmaxfield,nomura,neuenfeld}
in connection with the important recent advances in the black hole information paradox \cite{penington,aemm,islands,eastcoast,westcoast,micro} (whose applicability to massless gravity has been called into question in \cite{gengkarch,karchraju,rajunonsplitting}).

The second noteworthy feature is that, by construction, the state that we obtain in the single BCFT defined on the CFT causal diamond $D$ has exactly the same pattern of entanglement among its interior degrees of freedom as the vacuum $\ket{\Omega}$ of the CFT prior to Wilsonian integration over $D^c$. This implies that the dual bulk has exactly the same geometry as the original spacetime prior to introduction of the EOW brane, namely pure AdS.\footnote{As long as we ignore backreaction from the scalar field sourced by $\phuv$. As emphasized before, our procedure is not limited to pure AdS: it can be applied in any asymptotically locally AdS geometry.} It is in this context that an unambiguous match between our bulk and boundary Wilsonian reductions seems most plausible, given that both give access to precisely the same correlators. In particular, $\cE$ is exactly the region within which local insertions of $\phi(x,z)$ correspond to smeared versions of the dual operator $\cO(x)$ that are fully contained within $D$ \cite{dhw}. Our conclusion then is that (\ref{subregion3}) should be dual to the result of carrying out the integral over $\Phir$ in (\ref{DiamondInt2}), with the standard identification $J(x)=\phuv(x)$:
\begin{align}
\label{DiamondInt3}
Z_{\bcft}[\phuv]=
\int\mathcal{D}\Phi_{D} \, 
\exp\left(i(\icft[\Phi_{D}]
+\ib^{\mbox{\tiny N}}[\Phir]+\int_{D}\cO(x)\phuv(x))\right)~,
\end{align}
where $\Phir$ is understood to be determined by the Neumann-type edge condition analogous to (\ref{neumannir}).

\subsection{Extension to more general coarse-grainings} \label{extensionsubsec}

Both in the bulk and in the boundary, we can consider reductions to spacetime regions other than an entanglement wedge $\cE$ and its corresponding causal diamond $D$. For clarity, it will be important for us to distinguish between two types of reduction that are frequently taken to be synonymous: \emph{integrating out} vs.~\emph{tracing over}. In our usage here, 
the former phrase will refer to formulating the partition function of our system in terms of a path integral over some set of variables, and then carrying out some part of the integration, to be left with a \emph{reduced partition function}. On the other hand, we speak of tracing over when we have a state defined as a (possibly pure) density matrix on some time slice, choose some bipartitioning of the degrees of freedom into complementary subsets $A$ and $A\sc$, and take a sum of expectation values over a basis of states for $A\sc$ alone, to be left with a \emph{reduced state} that only refers to $A$. In familiar cases, this operation admits a path integral representation, and there is then a connection between the two concepts. A general result can be found in \cite{vijaymark}. But in such a context, the notion of integrating out is more general; in particular, it need not take place on a particular time slice.  

For our purposes, the main distinction between the two notions is the set of sources that we incorporate. When we integrate out, no particular set of sources is forced upon us, and if we choose to leave all sources turned on, the reduced partition function will allow us to determine \emph{all} correlators. By itself, then, the operation of integrating out does not necessarily amount to a loss of information: we could simply be making partial progress on the full calculation of interest. The situation is different when we trace over, because we have a clear bipartitioning of degrees of freedom, and when computing the partial trace, we must choose once and for all whether or not we have specific operators acting on $A\sc$. In short, we could say that sources involving $A\sc$ must be turned off, although it is more accurate to say that no differentiation with respect to such sources will be allowed, because that would amount to subtracting different reduced states.  

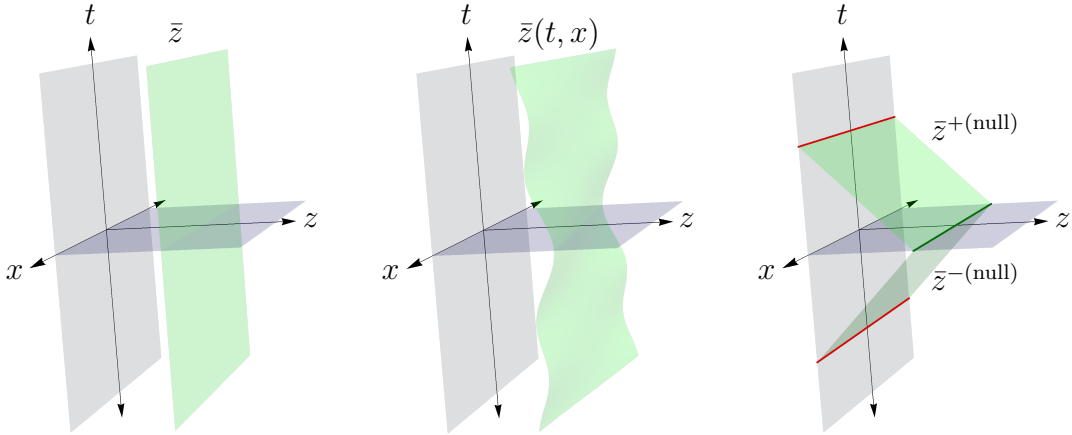
\begin{figure}[!t]
\begin{center}
  \includegraphics[width=5cm]{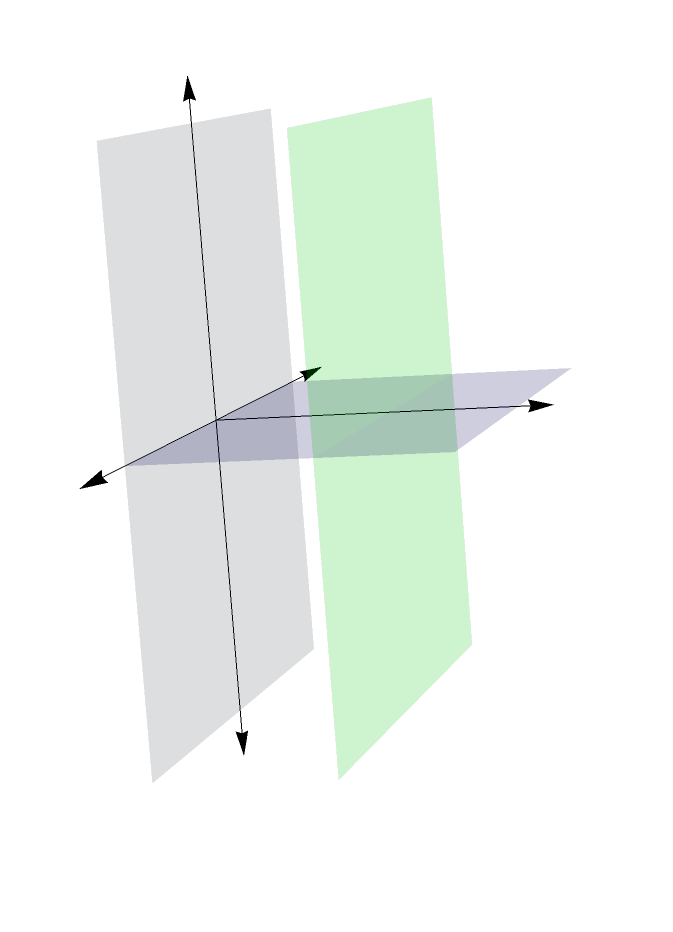}\includegraphics[width=5cm]{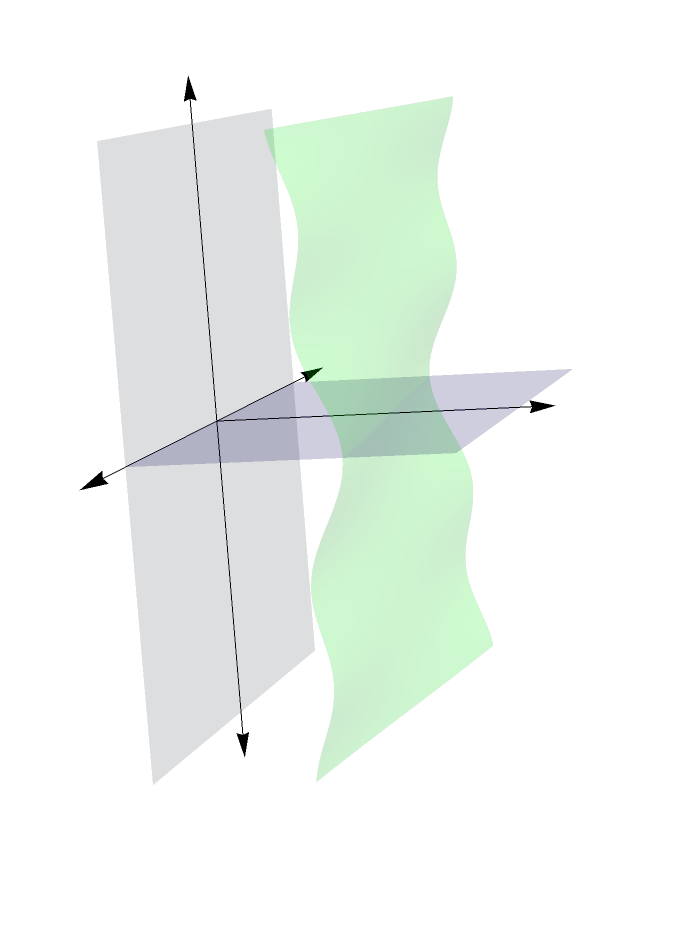}\includegraphics[width=5cm]{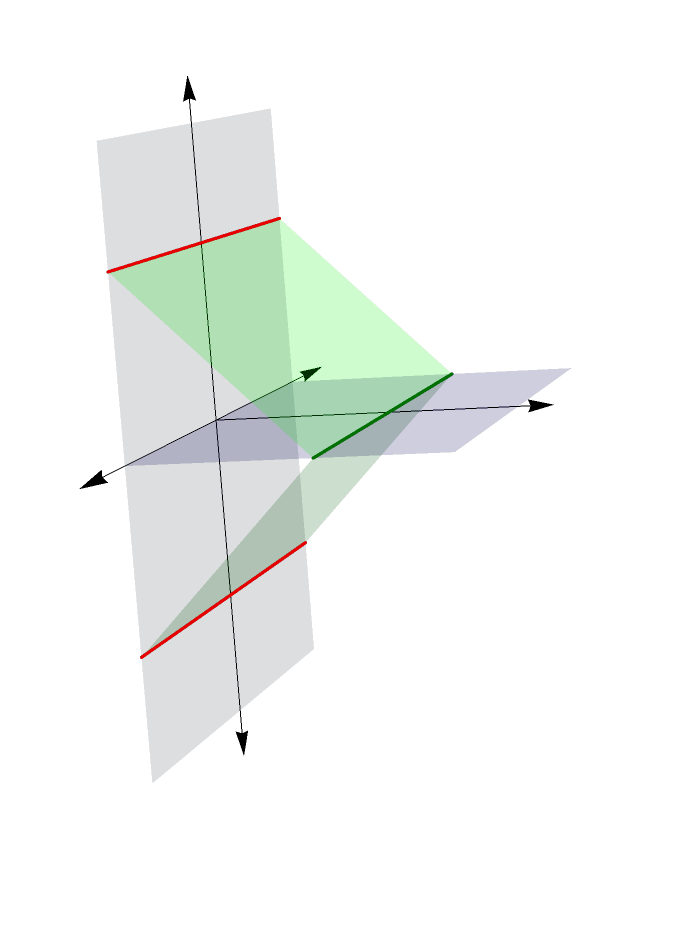}
  \hspace*{-1cm}
  \setlength{\unitlength}{1cm}
\begin{picture}(0,0)
\put(-14.15,3.15){$x$}
\put(-9.15,3.15){$x$}
\put(-4.15,3.15){$x$}
\put(-13.1,6.55){$t$}
\put(-8.1,6.55){$t$}
\put(-3.1,6.55){$t$}
\put(-10.2,3.84){$z$}
\put(-5.2,3.84){$z$}
\put(-0.2,3.84){$z$}
\put(-12,6.3){$\bar{z}$}
\put(-7.35,6.3){$\bar{z}(t,x)$}
\put(-1.85,5.0){$\bar{z}^{+\text{(null)}}$}
\put(-1.85,3.0){$\bar{z}^{-\text{(null)}}$}
\end{picture}
\end{center}
\vspace*{-0.8cm}
\caption{Left: vertical EOW brane, obtained by integrating out the IR portion of the geometry with $z>\bar{z}$ ($\forall\,\, t,x$). Center: undulating EOW brane, obtained by integrating out the IR portion of the geometry with $z>\bar{z}(t,x)$. Right: EOW brane that delineates a generalized entanglement wedge associated with a reduction in \emph{momentum} space, obtained by tracing over the IR modes ($z>\bar{z}$) of a state defined on a constant-$t$ slice. See the main text. The wedge is bounded by null surfaces $\bar{z}^{+\text{(null)}}$ and $\bar{z}^{-\text{(null)}}$, and covers a finite time strip on the boundary. 
\label{coarsegrainfig}}
\end{figure}

The considerations in this paper are completely general in the sense of integrating out. As explained in Section~\ref{spatialsubsec}, in the CFT we can reduce the partition function down to any spacetime region $R$, such as those exemplified in Fig.~\ref{reducedcftfig}. Likewise, in the bulk, our rememorization procedure can be employed to reduce to any given spacetime region $\cR$, such as those illustrated in Fig.~\ref{reducedbulkfig}.  
Additional examples of bulk IR coarse-grainings are shown in Fig.~\ref{coarsegrainfig}, to be discussed below. In all cases, the EOW brane action $\iir$ can be regarded as a functional of all possible sources, thereby keeping perfect memory of the physics in the region $\cR\sc$ that has been integrated out. 

Let us now move towards the tracing-over perspective, focusing first purely on the CFT side. A choice of spacetime region $R$ identifies the complement $R\sc$ that is  to  be  integrated  out.  But  what  are  the  corresponding  degrees of freedom in  a canonical presentation? One might think about different possibilities, obtained by cutting open the path integral in $R$ along some time slice, but there  is  in  fact a unique optimal answer.
Given  an open spacetime region $R$, its causal complement $R'$ is defined to be the largest open region that is spatially separated from $R$. Repeating this operation, we obtain $R''$, which is the smallest causally complete set that contains $R$. The important property of this causal completion is that the  algebra of operators in $R''$ coincides\footnote{Assuming Haag duality.} with that of $R$, meaning that all operators within $R''$ can be rewritten in terms of those in $R$--- see, e.g., \cite{wittenentanglement}. By construction, $R''$ is some causal diamond, that we will label $D$.
When we integrate out $R\sc$, we certainly retain the possibility of inserting arbitrary operators within $R$,  and consequently, within the larger region $D$. Choosing any time slice $\Sigma$ that contains the `waist' of $D$, we identify $A=\Sigma\cap D$. We then conclude that $A\sc=\Sigma\setminus A$ has been traced over:  no sources are allowed there, or indeed, in all of $D\sc$.  

If we wish, we can maintain the point of view that we have integrated out down to the original $R$, leaving the sources in $D\setminus R$ turned on. But it is more natural not to do so:  we would like to see explicitly within our reduced partition function all points where local operators can be reconstructed, i.e., all operators that commute with those in $A\sc$.  Relatedly, we would like to be able to cut open the given BCFT path integral to reveal the degrees of freedom  in all of $A$, and this would not be the case for an arbitrary $R$. In short, from the perspective of tracing over spatial degrees of freedom, there is no natural reason to consider a spacetime region that is not a causal diamond, $R\neq D$.

With this understanding, let us now discuss the bulk picture. Given a causal diamond $D$ in the CFT on which we allow ourselves to turn on sources,  from the integrating-out perspective we can rememorize the full partition function down to any bulk region $\cR$ that intersects the boundary on $D$. In other words, there are infinitely many EOW brane locations $\rir$ and corresponding IR actions $\iir$, in Neumann or Dirichlet presentation, that equally fulfill the goal of keeping memory of all correlators in $D$. They all define the same reduced state $\rho$. 
Within each time slice $\Sigma$, we could pass to a tensor network description
\cite{swingle,happy,nothappy,walltensor}, and the free choice of location of $\rir\cap\Sigma$ would then correspond to 
the ability of pushing the state 
along the network.\footnote{This ability in turn motivates the surface-state correspondence \cite{surfacestate}, and its more concrete incarnation in terms of path integral optimization \cite{pioptimization}. In a separate direction, it motivates the coarsening procedure put forth in \cite{nrs}.}
Back in the continuum, if each of these setups in Neumann form corresponds under AdS/BCFT to a specific BCFT edge action $\ib^N$, then all of these actions would also correctly define the same $\rho$. 
But with the same criterion as in the previous paragraph, we prefer to see explicitly in $\cR$ all bulk points where operator insertions can be reconstructed within $D$, so by subregion duality \cite{dhw}, we must choose $\cR=\cE$. 
%%Mention analog in terms of the tensor network: transformation is only unitary if we stop pushing at RT surface
We thus learn that, from the tracing-over perspective where we want to encode a specific state that has been reduced to some spatial region $A$, we are led uniquely to the rememorization procedure defined in Section~\ref{rememorizationsubsec}.  

For simplicity, up to this point we have been cavalier about one aspect\footnote{Mentioned briefly in footnote \ref{cutopenfootnote}.} that we now wish to make more precise. 
As we have just recalled, a very special feature of an entanglement wedge, prominent in the motivation for the present work and not shared by generic bulk regions, is its close relation with the reduced density matrix $\rho$ that is obtained by tracing over the degrees of freedom outside of a clearly identified spatial region $A$ in the boundary theory. The usual construction of $\rho$ in terms of path integration involves Euclidean segments that prepare
the ket and the bra for the desired state on the time slice $t=t_0$ where $A$ resides. 
Putting these segments together, one ends up with a path integral that has a narrow horizontal slit on $A$, and by specifying field profiles on the upper and lower lips of the slit, we obtain the matrix elements of $\rho$.
This density matrix then serves, as in (\ref{correlators}), to compute the expectation value of products of operators inserted at time $t=t_0$. Upon rotating back to Lorentzian signature, it equally determines the correlators of operators within the causal diamond $D$ of $A$  \cite{ch2}, because causality guarantees that these operators can be reexpressed as insertions on $A$. 
The extension in the CFT from $A$ to $D$ is mirrored in the bulk by the identification of the entanglement wedge $\cE$, which, as recalled in the Introduction, is likewise a causal domain of dependence determined by $A$ and its corresponding RT surface $\Gamma$. The entire story can be framed from the start in Lorentzian signature, in which case the contour of the path integral is of Schwinger-Keldysh type, doubling back in time to correctly account for the ket and the bra \cite{dlr}.

The point that is highlighted by the preceding paragraph is that, when we cut open the general kind of (bulk or boundary) path integral that we have been considering in this paper across some time slice $t=t_0$, we will only obtain a (total or reduced)  density matrix $\rho$ if the Hamiltonian and external sources happen to be invariant under time reversal about $t_0$.
Generically, the nontrivial time dependence implies that the bra and ket portions of the integral correspond to different states. What we have then is a \emph{transition matrix} $\tau$, precisely the setup that has recently been considered in \cite{pseudoentropy} in the Euclidean setting, where it gives rise to the notion of `pseudo-entropy'. Insertion of operators on $A$ or $D$ then yields a matrix element instead of an expectation value.  

Having clarified this, we can reconsider the case where in the CFT we integrate out down to an open spacetime region $R$ in the CFT that is \emph{not} a causal diamond. As explained in Sections~\ref{spatialsubsec}-\ref{dnsubsec}, this leads to a BCFT that is generally time-dependent, and should conceivably be dual to some specific bulk region $\cR$ capped off by an EOW brane with a suitable action $\iir$. Even though we know that we could naturally associate this setup with the causal completion $R''\supset R$ and its corresponding entanglement wedge $\cE\supset\cR$, if we wish, we could just take the boundary and bulk partition functions reduced to $R$ and $\cR$ at face value, using them only to compute correlators within the BCFT defined on $R$. In particular, if we cut the remaining path integral along some open spatial region $A$ whose causal diamond $D$ fits within $R$ and turn off sources in $D\sc$, the result will yield a transition matrix $\tau$ between two states in the BCFT.\footnote{Since by construction the boundary action of the BCFT encodes the spacetime region that has been integrated out, we could have situations where $\tau$ is in fact a reduced density matrix from the perspective of the original CFT (it is hermitian and has unit trace), but is reinterpreted as a transition matrix in the context of the BCFT.} It seems plausible that the entanglement wedge associated with $D$ would be fully contained within $\cR$.\footnote{Notice that here we are purposefully leaving out of consideration the case where $A$ is not open and includes some portion of the edge $\p R$ of the BCFT. In that case, the corresponding RT surface will of course end on the EOW brane at $\rir$ \cite{adsbcft}. Within our setup, inclusion of the edge indicates involvement of some of the CFT degrees of freedom that were originally in $R\sc$, so it is only natural that the corresponding entanglement wedge intersects $\rir$. The gravitational $\iir$ would account for the area of the portion of the RT surface that was originally in $\cR\sc$.}  The often studied case \cite{adsbcft} of EOW branes with constant negative tension provides examples where RT surfaces $\Gamma$ for some choices of $A$ in the CFT fail to be contained inside $\cR$, but the expectation then would be that the BCFTs produced by integrating out some spacetime region $R\sc$ are not of this kind.

This prompts us to ask whether there could be some direct interpretation, in terms of subregion duality, of a bulk region that is \emph{not} an entanglement wedge. In an AdS/BCFT scenario, the associated $\cR$ and $\iir$ by definition allow the correct determination of correlators in $R$.  
When $R$ is a causal diamond $D$, we know that the entanglement wedge $\cE$ is singled out as the bulk region that includes any and all spacetime points where local operators can be reconstructed as smeared CFT operators within $D$ \cite{dhw}. If it is indeed true that in the setup of the previous paragraph $\cR$ would contain all entanglement wedges whose diamonds fit within $R$, the bulk region in question would be some type of generalization of the usual notion of entanglement wedge for spacetime reductions. 

A different kind of generalization might exist for bulk regions such as those illustrated in Fig.~\ref{coarsegrainfig}. The first example depicts an EOW brane at a constant radial depth $\zir$, capping off a bulk region $\cR$, which is simply $z\le\zir$. The fact that the brane is not anchored on the boundary indicates that this is \emph{not} a spacetime reduction in the CFT. {}From the perspective of global AdS, the brane does end on the boundary, but it does so precisely on the edge of the Minkowski diamond $M$, which reinforces the inference that in this case $D$ is all of Minkowski spacetime. The corresponding entanglement wedge $\cE$ is of course all of the Poincar\'e wedge $\cP$, which evidently differs from $\cR$.\footnote{For large $\zir$, this choice of $\cR$ is a concrete example of an entanglement wedge $\cE$ regularized by a stretched horizon, as mentioned in footnote~\ref{stretchedhorizon}.}  
%%Mention somewhere rememorization from global AdS to Poincaré. Different boundary conditions at horizon encode different states.
%%Relatedly, rememorization of TFD.

As discussed around Eq.~(\ref{irintegration}), through the UV-IR connection \cite{uvir}, the standard field theory interpretation attached to  Fig.~\ref{coarsegrainfig} left is as a reduction in momentum space, where we integrate out, with some type of smooth cutoff, the IR modes 
$|\vec{p}|< 1/\zir$ at all $t$. 
The second example in the figure has the EOW brane at a variable depth $\zir(t,\vec{x})$, and as noted in the paragraph below (\ref{irintegration}), it is related to the previous case by a Weyl rescaling in the CFT. 

But in any of the examples of Fig.~\ref{coarsegrainfig}, we know that the mere process of integrating out does not automatically entail a restriction on the sources on which the resulting reduced partition function depends. It is only when we choose the sources that we obtain an association with a specific state (a $\rho$), or pair of states (a $\tau$), or family of states (a source-dependent $\rho$ or $\tau$).
Our choice of available sources feeds into the action and location of the EOW brane, and
determines which correlators are accessible to us within $R$.  Since $R\sc$ is empty in the first two examples of Fig.~\ref{coarsegrainfig} ($R=R''$ is the full Minkowski diamond $M$), if we are rememorizing the original full state (e.g., the CFT vacuum), standard subregion duality makes it natural to allow ourselves all sources, thereby obtaining all  correlators. This is what we discussed above: $\cR$ would be naturally enlarged to the full Poincar\'e wedge $\cP$. 

There are two separate routes we could follow to depart from this conclusion. The first is to turn off some set of sources defined not in position but in momentum space. E.g., in Fig.~\ref{coarsegrainfig} left, the UV-IR connection suggests that we turn off all Fourier modes with $\vec{p}<1/\bar{z}$ at all times. Correlators of strictly local operators would then be unavailable. As mentioned before, the association between radial position and a momentum scale is not sharp \cite{grozdanov,kiritsis}, so the appropriate sense of `turning off' should likewise not be abrupt. 
Whatever the association, in Fig.~\ref{coarsegrainfig} center we would allow only those sources that are obtained from those of the left figure via the corresponding Weyl transformation.  

Having shifted attention from position to momentum space, we can think of preparing a CFT state at a given time and tracing over a certain set of momentum modes, to define a genuine reduced density matrix or transition matrix, as in \cite{vijaymark}. In generic field theories, there would be no obvious automatic implication about other times; but holographic QFTs have an emergent causal structure in momentum space,
made evident in the bulk. Guided by this, we could propose an extension of the concept of entanglement wedge to momentum space reductions. The simplest example is shown in Fig.~\ref{coarsegrainfig} right: upon tracing over the  IR modes $|\vec{p}|<1/\bar{z}$ at time $t=t_0$,  we identify the bulk codimension-2 surface $\Pi$ located at $(t=t_0, z=\bar{z})$ as the analog of the standard RT surface,
and define our region $\cR$ of interest as the bulk domain of dependence of any spacelike codimension-1 surface extending from $\Pi$ to the boundary slice at $t=t_0$. 

The way in which the corresponding EOW brane intersects the boundary indicates that, starting from this reduced specification of the initial state, only a finite time window is available in the CFT. Importantly, we should not take this to mean that we can compute all correlators of \emph{local} operators within the time strip $R$, because the causal completion $R''$ is all of Minkowski space, and then we would be back in the scenario we wanted to depart from. Rather, through the UV-IR connection, causality in the bulk suggests that the set $\tilde{J}(t,\vec{p})$ of sources available to us is more and more restricted to the UV as we move further away from the $t=t_0$ slice, until there comes a time where no sources whatsoever are allowed.    

Notice that this perspective is markedly different from the one adopted in the hole-ographic approach \cite{hole-ography,myers,cds,veronikaresidual,hmw,nutsandbolts,bartekinformation,integralgeometry,whatsthepoint,br,bartekyaithd,stereoscopy,diamondography}. 
In that method, one begins by identifying the set of RT surfaces $\Gamma(\lambda)$ tangent\footnote{The tangency condition can be relaxed somewhat, using the condition of `null vector alignment' discovered in \cite{hmw}.} to the surface $\Pi$, and then relies on subregion duality to translate this into a continuous family of reduced density matrices $\rho(\lambda)$ in the boundary theory. This approach is equally applicable to more general codimension-2 surfaces $\gamma$ situated at some location $\bar{z}(x)$, including those that only sweep over a finite spatial extension in the boundary.  In all cases, one is led to a \emph{collection} of density matrices, each of which refers to a specific \emph{spatial} reduction in the CFT, and the area of $\gamma$ is computed by the `differential entropy' of $\{\rho(\lambda)\}$.\footnote{A somewhat similar approach for losing the IR is that of \cite{eflosing}, and \cite{nrs} also contemplates a collection of density matrices, with the opposite aim of losing the UV.} 
In the present paper we are envisioning instead a \emph{single} density matrix resulting from \emph{momentum} space reduction, as originally imagined in \cite{vijaymark}.\footnote{And
somewhat like \cite{ewcoarse}, but with a different choice of allowed measurements.} 

The two setups are sharply distinguished by the type of correlators one has access to, and by construction, our reduced partition function has entanglement information that is unavailable in hole-ography. 
Physically, the difference is that the latter approach considers a continuous family of local observers, each with full information about her own causal diamond, finitely extended in time, but with the proviso that different observers are \emph{not} allowed to make joint measurements. 
Mathematically, this means that altogether one is focusing not on an algebra, but merely on a set of operators, along the lines of the general information-theoretic analysis of \cite{donald}. A precise protocol associated with differential entropy was identified in \cite{bartekinformation}. In our approach,  joint measurements are allowed, so we do remain within the more familiar setting of an operator algebra, but our `observers' make measurements (somewhat) localized in momentum space instead of position space.    
One should be able to treat in an analogous fashion situations where the set $P$ of momentum modes that remain after reduction is different. For instance, $P$ could comprise two or more bands of values of $|\vec{p}|$, and again the resulting domain of dependence $\cR$ in the bulk would be interpreted as a restriction to the algebra of operators that commute with those in $P^c$. 
%%Go back to momentum wedge $\cP$. It is causally complete, and we can consider its causal complement $\cP'$. On a time slice, they define sets $P$ and $P\sc$. The point then is that we could restrict sources to be only those that commute with $P\sc$. By this we do not mean some vague Fourier mode prescription, but local bulk operators within $P\sc$. This does seem to be the right notion. Now we \emph{do not} have full acess to all $\cO(x)$ in the time strip, so we do not reconstruct all of $M$. This is what is special about the momentum wedge: in the other 2 pictures, causal completion in the bulk leads one to all of the Poincar\'e wedge. Notice nesting is automatic.

A second route to depart from the familiar $R\!=\!M\;\leftrightarrow\;\cR\!=\!\cP$ identification is to use the fact that, for a given choice of $\cR$, we can sweep over all possible overall states by considering all possible EOW brane actions $\iir$. Among these options, there would be one that is appropriate for instance for outright (sharp or smooth) \emph{truncation} of the IR modes.  This is distinct from the restriction on the sources discussed in the preceding paragraphs, because Wilsonian integration by itself does not truncate, it rememorizes.\footnote{The main lessons here about IR rememorization apply as well in the usual case where we are moving the UV cutoff surface inward \cite{hp,flr,bgl}. There is a clear distinction between holographic Wilsonian renormalization and truncating. In the former case, independently of where we place the UV cutoff surface $\underline{\p\cR}$, we would still have access to \emph{all} correlators if we allow ourselves arbitrary sources.}
A different example in the same spirit is to take $\cR$ to be a standard entanglement wedge $\cE$, and then scan over the possible $\iir$ to find the one that defines the state that is `optimal' in the sense of entanglement of purification \cite{eop}, as implemented holographically in \cite{takayanagi,phuc} (see also \cite{bao2,rindler,umemoto,kudlerflamryu,tamaoka,duttafaulkner,caputa}).
We leave to future work the detailed exploration of these various applications and generalizations of our rememorization method.

\section{Rememorization of generic scalar correlators} \label{scalarsec}

A standard scenario where we can illustrate our general rememorization procedure is in the computation of correlators of a single scalar operator $\cO(x)$ in a large-$c$ CFT$_d$ on Minkowski spacetime, which is dual to a free scalar field $\phi(x,z)$ on Poincar\'e AdS$_{d+1}$, Eq.~(\ref{poincare}). As recalled in the Introduction, the GKPW recipe \cite{gkp,w} equates the partition functions on both sides as in (\ref{zz}), identifying the external source $J(x)$ of $\mathcal{O}(x)$ in the CFT with the asymptotic boundary condition $\phuv(x)$ of the dual field. 

Working in Euclidean signature, the bulk partition function is given by
\begin{align}
\label{zgrav0001}
   \zgrav[\phuv] 
 &=\int_{\phi(x,\epsilon) =\epsilon^{d-\Delta}\phuv(x)} 
\hspace*{-2cm}\mathcal{D}\phi\,  
 \exp\left(-\ibulk[\phi]
 -\iuv[\phuv]\right)~,
\end{align}
where $z=\epsilon$ is the UV radial cutoff, and $\Delta$ denotes the scaling dimension of $\cO(x)$.
The bulk action is
\begin{equation}
\label{eqn:secSF01}
      \ibulk[\phi]=\frac{1}{2}\int d^{d+1} x \sqrt{g}\left[g^{mn}\partial_{m}\phi\partial_{n}\phi +M^2 \phi^2 \right]~,
\end{equation}
with 
$M^2 L^2 =\Delta(\Delta-d)$. As always, $\iuv[\phuv]$ denotes the counterterm boundary action for holographic renormalization
\cite{dhss,skenderis}, whose explicit form will not be needed here. 

\subsection{A simple example: wall at constant~$z$} \label{wallsubsec}

The most obvious way to separate the bulk spacetime is by means of a wall at $z=\zir$. See Fig.~\ref{coarsegrainfig} left. The regions to the left and to the right of the wall correspond respectively to the UV and IR of the CFT. Contrary to the standard Wilsonian elimination of the UV region, here we want to integrate out the IR component. This reduction will generate a contribution to the effective action of the remaining geometry that encodes the information of the IR. 

Specifically, in analogy with (\ref{iraction}) we need to calculate 
\begin{align}
\label{effective0003}
    \exp(-\iir[\phir])=
    \int_{\phi(x,\zir) =\phir(x)} \hspace*{-1.5cm}\mathcal{D}\phi_{z>\zir}
    \,\exp\left(-\ibulk[\phi]\right),
\end{align}
where $\phir(x)$ is the boundary condition for the fields on the resulting EOW brane at $\zir$. Since the path integral is quadratic, the saddle-point approximation is exact. 
For the on-shell evaluation of (\ref{effective0003}), we need the classical solution of the KG equation
in the (Wick-rotated version of the) metric (\ref{poincare}), 
\begin{equation}
\label{eqn:secSF04}
    z^{d+1} \partial_{z} \left(\frac{\partial_{z}\phi}{z^{d-1}} \right)+z^2 \delta^{\mu \nu}\partial_{\mu}\phi\partial_{\nu}\phi-M^2 L^2 \phi=0~.
\end{equation}
As usual, due to translational symmetry, it is useful to decompose into Fourier modes,
$\phi(x,z)=\int \frac{d^{d} p}{(2\pi)^d} \,\tilde{\phi}(p,z)e^{i p\cdot x}$. 
The general solution is then found to be \cite{gkp,w} 
\begin{align}
\label{eqn:secSF05}
      \tilde{\phi}(p,z)= z^{\frac{d}{2}}\left(C_{1}(p)I_{\Delta- \frac{d}{2}}(pz)+C_{2}(p)K_{\Delta- \frac{d}{2}}(pz)\right)~.
\end{align}
The coefficients $C_{1}(p)$ and $C_{2}(p)$ take specific forms depending on the boundary conditions imposed. For the standard time-ordered correlators, we impose regularity as $z\rightarrow \infty$, which sets $C_1(p)=0\;\,\forall\,p$. At the wall, we must enforce the Dirichlet condition  
\begin{align}
        \phi(p,\zir) &=\tilde{\phir}(p)~,
\quad\mbox{with}\quad
\tilde{\phir}(p)
        \equiv\int d^d x\,
        \phir(x)e^{-i p\cdot x}~.
\end{align}
This determines $C_2(p)$, singling out the momentum-space solution
\begin{equation}
     \tilde{\phi}^{\text{cl}}(p,z)= \left(\frac{z}{\zir}\right)^{\frac{d}{2}} \frac{K_{\Delta-\frac{d}{2}}(pz)}
     {K_{\Delta-\frac{d}{2}}(p\zir)}
     \tilde{\phir}(p)~.
\end{equation}
With this solution, we can evaluate the on-shell action. The result can be expressed as a surface term at $z=\zir$,
\begin{align}
\label{IRBterm}
    \iir[\phir] &=\frac{1}{2}\int_{\partial \cR}d^{d} x \sqrt{h}\, n^{m} \phi^{\text{cl}} \partial_{m} \phi^{\text{cl}}~,
\end{align}
with $h$ the induced metric and $n$ the outward unit normal. In momentum space, this is
\begin{align}
    \iir[\phir]&=
          -\frac{L^{d-1}}{2\zir^{d-1}}\int\frac{d^d p}{(2\pi)^d}  \left.\Tilde{\phi}^{\text{cl}}(-p,z) \partial_{z} \Tilde{\phi}^{\text{cl}}(p,z) \right|_{z=\zir}~.
\end{align}
The derivative in the integrand can be evaluated explicitly, obtaining \begin{align}
     \left.
     \partial_{z} \Tilde{\phi}^{\text{cl}} \right|_{z=\zir} &=  \left.\partial_{z}\left(  \left(\frac{z}{\zir}\right)^{\frac{d}{2}}\frac{K_{\Delta-\frac{d}{2}}(pz)}{K_{\Delta-\frac{d}{2}}(p\zir)} \right)\right|_{z=\zir}
     \tilde{\phir}(p)~.
\end{align}
At this point, it is convenient to define 
\begin{align}
\label{Boperator}
    \hat{\mathcal{W}}\phir(x)
    \equiv-\frac{\bar{z}}{L}\int \frac{d^d p}{(2\pi)^d}  \left.\partial_{z}\left(  \left(\frac{z}{\zir}\right)^{\frac{d}{2}}
    \frac{K_{\Delta-\frac{d}{2}}(pz)}{K_{\Delta-\frac{d}{2}}(p\zir)} \right)\right|_{z=\zir}\tilde{\phir}(p)
    e^{ip\cdot x}~. 
\end{align}
%%locality at small p
Using this in (\ref{IRBterm}),  we are left with
\begin{align}
\label{BDRYterm}
   \iir[\phir]=\frac{1}{2}
   \int_{z=\zir} d^{d} x\sqrt{h}\,\phir(x)
   \hat{\mathcal{W}}\phir(x)~.
\end{align}
Note that this is a nonlocal expression, because, as seen in (\ref{Boperator}), the operator $\hat{\mathcal{W}}$ acts on $\phir(x)$ with an infinite series of spacetime derivatives. 

The boundary term (\ref{BDRYterm}) encodes the information of the classical profile of the scalar field in the IR region of the bulk, $z>\zir$. As explained in the previous section, this term supplements the action in the UV region, and determines the suitable boundary condition for the field $\phi$ at $z=\zir$. This guarantees the coincidence between correlators computed with the reduced geometry and those obtained from the complete spacetime. 

To see this explicitly, notice that variation of the effective action for the UV region,
$ I\eff\equiv\ibulk[\phi]
    +\iir[\phir]+\iuv[\phuv]$, yields the following boundary terms at $z=\zir$:
\begin{align}
\label{variationB}
    \delta I\eff\supset
    \int_{z=\zir}d^{d}x \sqrt{h} n^{m}\partial_{m}\phi\delta\phi  +\frac{1}{2}\int_{z=\zir} d^{d} x\sqrt{h}\delta\phi\hat{\mathcal{W}}\phi
    + \frac{1}{2}\int_{z=\zir} d^{d} x\sqrt{h}\phi\hat{\mathcal{W}}\delta\phi~.
\end{align}
We then need to integrate by parts the final term, but the nonlocal nature of the operator $\hat{\mathcal{W}}$ makes this a bit nontrivial in the position-space representation. We will return to this point in the next subsection. Here, it is simplest to change to the momentum-space representation, $\phi(x,z)=\int \frac{d^{d} p}{(2\pi)^d} \tilde{\phi}(p,z) e^{i p\cdot x}$.
Using (\ref{Boperator}), we see then that the vanishing of (\ref{variationB}) implies that one of the following two boundary conditions must hold:
\begin{align}
\label{dirichletwall}
\mbox{Dirichlet (IR)}\qquad\qquad\;\;\;
\tilde{\phi}(p,z)|_{z=\bar{z}}&=\tilde{\phir}(p)~,
\\
\label{neumannwall}
\mbox{Neumann (IR)}\qquad\;\;\;\;\,
\left.\partial_{z}\tilde{\phi}(p,z)\right|_{z=\zir}
&= \left.\partial_{z}\left(  \left(\frac{z}{\zir}\right)^{\frac{d}{2}}
\frac{K_{\Delta-\frac{d}{2}}(pz)}{K_{\Delta-\frac{d}{2}}(p\zir)} \right)\right|_{z=\zir}\tilde{\phi}(p,\bar{z})~.
\end{align}
As explained in Section~\ref{dnsubsec}, these two alternatives are appropriate before and after performing the $\cD\phir$ path integral, respectively. Part of the content of that section was a proof of the equivalence between the Dirichlet and Neumann approaches for the case of a free bulk scalar field on a general background, so in our pure AdS$_{d+1}$ analysis here, we will focus on illustrating the Neumann approach, which is the more efficient of the two.  Our task is then to enforce (\ref{neumannwall}) at the wall, together with the standard Dirichlet condition at the AdS boundary,
\begin{align}
\label{dirichletboundary}
\mbox{Dirichlet (UV)}\qquad\qquad\;\;\;
\tilde{\phi}(p,z)|_{z=\epsilon}&=\epsilon^{d-\Delta}\tilde{\phuv}(p)~.
\end{align}

A general momentum-space solution to (\ref{eqn:secSF04}) is given by (\ref{eqn:secSF05}). Applying it in conjunction with (\ref{neumannwall}) and (\ref{dirichletboundary}), we deduce that
%%double-check
\begin{align}
      \phi(x,z)=\int \frac{d^{d} p}{(2\pi)^d}  \left(\frac{z}{\epsilon}\right)^{\frac{d}{2}}\frac{K_{\Delta-\frac{d}{2}}(pz)}{K_{\Delta-\frac{d}{2}}(p\epsilon)}\tilde{\phuv}(p) e^{ip\cdot x}~,
\end{align}
which as expected, is exactly equal to the solution in the entire spacetime with the same boundary condition at $z=\epsilon$ and the regularity  condition at the Poincar\'e horizon. 
This result confirms that the role of the boundary term, obtained via the Wilsonian reduction of the IR, is to keep the memory of the information of that region, by supplying the correct boundary conditions at the interface. 

\subsection{General recipe}\label{generalsubsec}

In the preceding subsection, we performed a reduction to a region of the spacetime delimited by a wall at $z=\zir$ (which becomes then the location of the EOW brane that arises from the reduction). In that example, translational symmetry greatly simplifies the task of finding an explicit solution of the Klein-Gordon equation with the prescribed boundary conditions. Nevertheless, our rememorization procedure can be applied to more general regions in the bulk. The example that was the initial motivation for this work is the entanglement wedge $\cE$ associated with a spatial region $A$ in the CFT, described in Section~\ref{wedgesubsubsec}. In that context, the reduction is performed over the exterior of the entanglement wedge, $\cE^{\mathsf{c}}$. 
In Section~\ref{extensionsubsec} we have discussed various generalizations. 

In the most general case, one reduces to some spacetime region $\cR$ in the bulk of an asymptotically locally AdS spacetime, which delineates some spacetime region $R$ in the boundary.
At large $c$, the saddle point evaluation translates into a linear PDE problem with arbitrary Dirichlet boundary conditions on the specified interface. 

In more detail, within $\cR^{\mathsf{c}}$ we ought to solve
\begin{align}
\label{KGphi}
\left(\DAlambert-M^2\right)\phi(x^m)=0~,
\end{align}
with boundary condition
\begin{align}
    \phi(x,z)=\phir(x) \qquad \text{if} \qquad (x,z)\in \rir~, 
\end{align}
where $\rir$ denotes the interface between $\cR$ and $\cR^{\mathsf{c}}$ (the eventual location of the EOW brane), and $\phir$ is a specific but arbitrary profile. If the boundary of $\cR^{\mathsf{c}}$ has other components aside from $\rir$, appropriate boundary conditions must be specified there as well. For instance, if $\cR=\cE$ is the entanglement wedge in Poincar\'e AdS associated with a causal diamond $D$ in the CFT, then we must turn off the source $J(x)$ in $D^{\mathsf{c}}$, and pick boundary conditions at the Poincar\'e horizon (e.g., regularity in the Euclidean description, which is appropriate if we wish to compute the time-ordered correlators). 

We can obtain a general solution to the problem if we first tackle the problem of finding the associated Green's function 
\begin{align}
\left(\DAlambert-M^2\right)G(x^m,x'^m)=\frac{1}{\sqrt{\abs{g}}}\delta(x^m-x'^m)~,
\end{align}
with the requirement that
\begin{align}\label{greensbc}
    G(x^m,x'^m)=0 \qquad \text{if} \qquad x^m\in \p(\cR^{\mathsf{c}})~.
\end{align}
In the following, we will specify the prescription to compute $\phi(x^m)$ given $G(x^m,x'^m)$. 

With the help of Green's identities, we can prove the following modification that incorporates the Klein-Gordon operator \cite{haberman}:
\begin{align}
\label{KGidentity}
    \int_{\mathcal{M}} d^{D}x' \sqrt{\abs{g}} \left\{v\left(\DAlambert-M^2\right)u- u\left(\DAlambert-M^2\right)v \right\} &= \nonumber \\
   & \int_{\partial \mathcal{M}} d^{D-1}x' \sqrt{\abs{h}} \left(v n^{m} \partial_{m} u -u n^{m} \partial_{m} v \right),
\end{align}
where $u$ and $v$ are two smooth functions defined in $\mathcal{M}$, $h_{ab}$ is the induced metric on the boundary $\partial \mathcal{M}$ and $n^{m}$ is the outward-pointing normal to $\partial \mathcal{M}$. The previous expression is stated in general notation, but for our purposes, we identify $\mathcal{M}$ with $\cR\sc$, and set the dimension to $d+1$. If we replace 
$u=\phi$ and $v=G$,
then Eq.~(\ref{KGidentity})  simplifies enormously. The first term in the left-hand side gives zero contribution since $\phi$ is assumed to be a solution to (\ref{KGphi}), while the second term gives the value of the field at $x'$. In the right-hand side, the boundary conditions of the Green's function make the first term vanish.
Therefore, we 
obtain\footnote{For brevity, we assume here that $\phi$ is meant to have vanishing boundary conditions on $\p(\cR^{\mathsf{c}})\setminus\rir$. Were this not the case, there would be an additional, $\phir$-independent term in (\ref{eqnIRtermGene}). This would not change anything essential in what follows.}
\begin{align}
\label{eqnIRtermGene}
    \phi(x^m)=\int_{\rir} d^{d}x' \sqrt{h(x')}\, \phir(x') n^{m} \partial_{m} G (x',x)~.
\end{align}
This expression is convenient, since its dependence on the boundary condition is manifest, and the evaluation of the Green's function can at least be carried out numerically. 

Substitution of (\ref{eqnIRtermGene}) in (\ref{IRBterm}) gives 
\begin{align}
    \iir[\phir]=\int_{\rir} d^{d}x  d^{d}x' \sqrt{h(x)}\sqrt{h(x')} n^{l}(x) n^{m}(x') \partial_{l}\partial_{m} G (x,x') \phir(x)\phir(x').
\end{align}
Note here that the crucial information of the region that has been integrated out, $\cR^{\mathsf{c}}$, is encoded in the Green's function $G$, because the profile $\phir$ is arbitrary. The resulting effective action in $\cR$ is thus
\begin{align}
    I_{\text{eff}}^{\cR}
    =I_{\text{bulk}}[\phi_{\cR}]
    +\iir[\phir]+\iuv[\phuv]~.
\end{align}

Consider now the variational principle based on 
$ I^{\cR}_{\text{eff}}$~. As usual, variation of the bulk term gives rise to a term that is proportional to the EOM and a term that is related to the normal derivative of the field at the interface. Variation of  \iir\  gives 
\begin{align}
     \delta \iir[\phir]
     &=\int_{\rir} d^{d}x  d^{d}x' \sqrt{h(x)}\sqrt{h(x')} n^{l}(x) n^{m}(x')\left[ \partial_{l}\partial_{m} G (x,x') \right. \nonumber\\
    & \qquad \qquad \qquad \qquad \qquad \qquad \qquad \qquad \qquad \left. + \partial_{l}\partial_{m} G (x',x)    \right] \phir(x') \delta\phir(x)~.
\end{align}
Based on the form of this expression, it is convenient to define the nonlocal operator 
\begin{align}\label{w}
    \hat{\mathcal{W}}\phi(x)
    \equiv\int_{\rir} d^{d}x' \sqrt{h(x')}n^{l}(x)n^{m}(x') \left[ \partial_{l}\partial_{m} G (x,x') + \partial_{l}\partial_{m} G (x',x)    \right] \phi(x')~.
\end{align}
%%Express IR action directly in terms of W?
Using this, we can rewrite the variation of the infrared action as
\begin{align} \label{deltair}
    \delta I_{\text{IR}}[\phir]&= \int_{\partial \mathcal{E}_{A}} d^{d}x \sqrt{h}  \hat{\mathcal{W}}\phir\delta\phir~.
\end{align}

The boundary variation (\ref{deltair}), supplemented with the variation of the bulk term, gives rise again to (\ref{variationB}), now with a more general definition for the operator $\hat{\mathcal{W}}$. {}From this line of reasoning, we see then that $\hat{\mathcal{W}}$ is a key piece of information of the reduction. In the end, we deduce the following two alternative boundary conditions for the field:
\begin{eqnarray}
{}&{}&\mbox{Dirichlet (IR)}\qquad\qquad\;\;\;
\phi(x)|_{\rir
}=\phir(x)~,
\label{dirichletgral}
\\
{}&{}&\mbox{Neumann (IR)}\qquad
n^m\p_m\phi(x)|_{\rir}
=\hat{\mathcal{W}}\phi(x)~.
\label{neumanngral}
\end{eqnarray}
As discussed in Section~\ref{dnsubsec}, the former is called for prior to carrying out the $\cD\phir$ path integral, and the latter is enforced as a result of that integral. Eq.~(\ref{neumanngral}) is a more explicit form of (\ref{neumann}): through (\ref{w}), the determination of the appropriate Neumann boundary condition and the resulting IR action for general $\cR$ has been reduced to finding the Green's function $G$ satisfying (\ref{greensbc}). The compact form of our notation should not obscure the fact that (\ref{neumanngral}) is a \emph{nonlocal} boundary condition, because $\hat{\mathcal{W}}$ denotes the convolution (\ref{w}). 

The geometric interpretation of (\ref{neumanngral}) is as follows. In the saddle-point approximation, the  full partition function (\ref{zgrav0001}) is entirely determined by the solution $\phi^{\mbox{\tiny cl}}$ that interpolates between the prescribed UV boundary condition $\phuv(x)$ and the appropriate IR condition (e.g., in Euclidean AdS, regularity at what would have been the Poincar\'e horizon). After rememorizing, the role of \iir\ is to pick out the \emph{same} solution $\phi^{\mbox{\tiny cl}}$ purely within $\cR$. Across $\rir$, this solution is continuous, and its normal derivative is also continuous. But when we split the path integral into the portions on $\cR$ and $\cR^{\mathsf{c}}$, and prior to carrying out the integral over $\phir$, the normal derivatives of the inner and outer saddles do not match. The net result of the $\cD\phir$ integral is to enforce the boundary condition (\ref{neumanngral}), which is precisely the statement that the two normal derivatives match. 
%%Point to some eqn for this. At least in previous section.
This ensures that we have found
$\phi^{\mbox{\tiny cl}}$. Since this solution was originally picked out by standard boundary conditions at the IR and UV ends of the full geometry, it naturally  does not satisfy any simple requirement on the intermediate surface $\rir$. This explains  the nonlocal nature of (\ref{neumanngral}): the correct normal derivative at any given point depends on the value of $\phir$ all over $\rir$.

\section{Correlators at large $\Delta$: geodesic approximation} \label{geodesicsec}

As noted above, for massive scalar fields, a near-boundary analysis reveals that the mass $M$ of the bulk field  must be related to the scaling dimension of the dual operator $\Delta$ according to $M^2L^2=\Delta(\Delta-d)$. For heavy fields, $ML\gg1$, the above condition implies that we are dealing with operators with large $\Delta\gg d$. In this limit one can make use of the saddle-point approximation
\cite{balasubramanianross,louko}, replacing the path integral over field configurations with a path integral over trajectories in a first-quantization language, and then considering the most relevant saddle. The key point is that, in this approximation, the correlators become localized over classical trajectories, i.e., geodesics, which in turn can be understood and visualized in a purely geometric way.

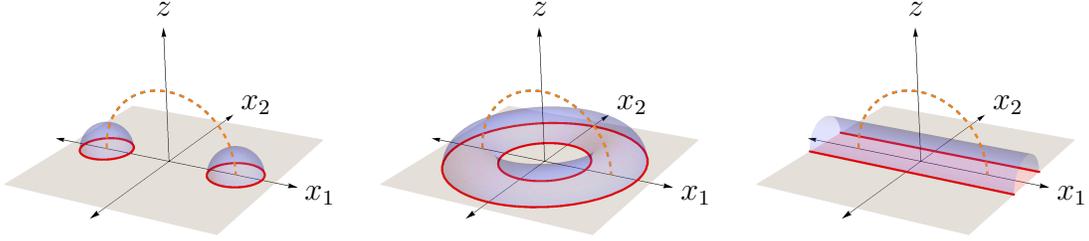
\begin{figure}[!t]
\begin{center}
  \hspace{-10mm}\includegraphics[width=5.5cm,trim={5mm 5mm 5mm 5mm},clip]{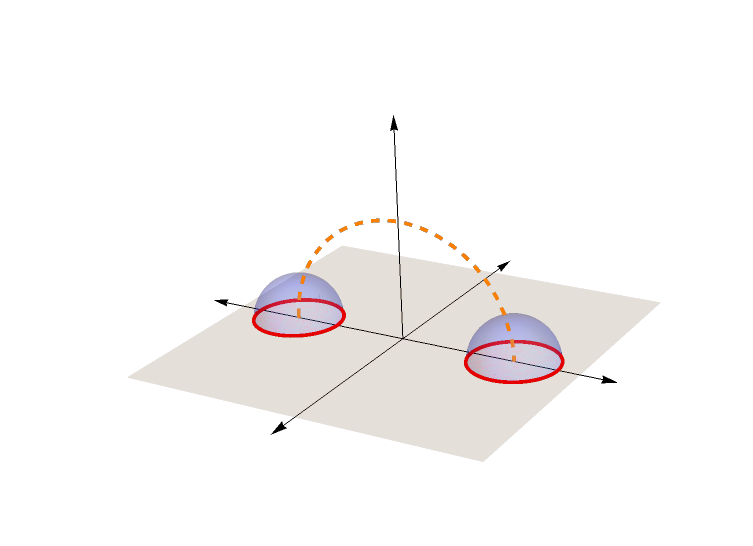}\hspace{-5mm}\includegraphics[width=5.5cm,trim={5mm 5mm 5mm 5mm},clip]{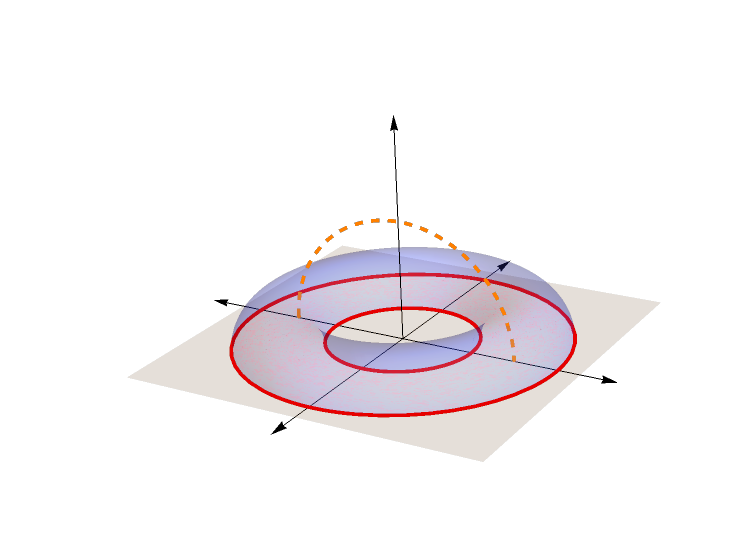}\hspace{-5mm}\includegraphics[width=5.5cm,trim={5mm 5mm 5mm 5mm},clip]{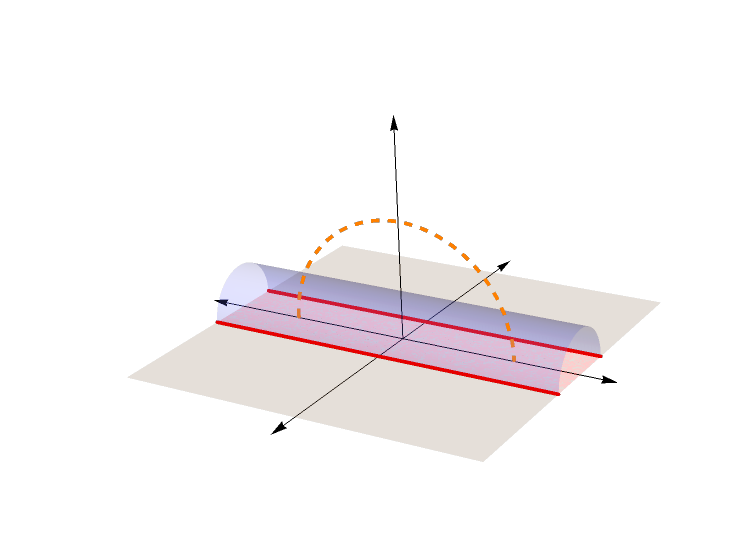}
  \hspace*{-1cm}
  \setlength{\unitlength}{1cm}
\begin{picture}(0,0)
\put(-12.13,3.35){$z$}
\put(-7.13,3.35){$z$}
\put(-2.13,3.35){$z$}
\put(-11.0,2.1){$x_2$}
\put(-6.0,2.1){$x_2$}
\put(-1.0,2.1){$x_2$}
\put(-10.15,0.9){$x_1$}
\put(-5.15,0.9){$x_1$}
\put(-0.15,0.9){$x_1$}
\end{picture}
\end{center}
\vspace*{-0.8cm}
\caption{Examples of geodesics (dotted orange) anchored on some region $A$ (light red) that exit their corresponding entanglement wedges, with $A$ being two widely separated spheres (left), an annulus (center), and an infinite strip (right). In all three cases, we have shown only the part of the entanglement wedges that lies on the time slice where the chosen region $A$ resides.
\label{geoexamplesfig}}
\end{figure}

Throughout this paper, we have emphasized that, to compute correlators accessible through a reduced density matrix $\rho$ as in (\ref{correlators}), the GKPW prescription requires knowledge of the profile of the bulk field $\phi(x,z)$ beyond the corresponding entanglement wedge $\cE$. In the large $\Delta$ limit where the calculation can be reformulated in terms of geodesics, the analogous statement is that the relevant geodesics will generally exit $\cE$. We illustrate this with some concrete examples in Fig.~\ref{geoexamplesfig}. 
The implication is the same as in the previous sections: for subregion duality to hold, it must be true that the information about the portion of the geodesics that probes $\cE\sc$ can be completely encoded in the infrared boundary term that we end up with after we have carried out the rememorization process defined in Section~\ref{wedgesubsubsec}. The geodesic approximation is, therefore, a natural arena to analyze this problem in a quantitative fashion. 
In this section, we carry out this investigation in a couple of concrete examples. We also take the opportunity to illustrate the equivalence, discussed in \S~\ref{dnsubsec}, between the Dirichlet and Neunmann approaches.

Before moving on to the examples to be worked out, we must discuss in more detail
the basic tool that we will be using in this section, namely, the calculation of correlators using first-quantized language.
 Using the BDHM or `extrapolate' dictionary \cite{bdhm,bkextrapolate,hs}, we can relate the CFT $n$-point function to the bulk $n$-point function with points taken close to the boundary of AdS,
\begin{align}
\label{eqn:BDHM}
\langle\cO(x_1)\cdots\cO(x_n)\rangle_{\cft}
=\lim_{z\rightarrow 0} z^{-n\Delta}
\expval{\phi(x_1,z)\cdots\phi(x_n,z)}_{\bulk}~,
\end{align}
with $\Delta$ is the scaling dimension of the operator $\cO$. In the large central charge limit, $c\to\infty$, the bulk field $\phi$ becomes free, so its $n$-point correlators vanish for odd $n$ and factorize into $2$-point functions for even $n$. At large but finite $c$, interactions can be treated perturbatively, in terms of the standard Witten diagrams, constructed with vertices and propagators. For simplicity, we will restrict our analysis to the free case.

We then formulate the calculation of the free bulk-to-bulk propagator not as a path integral over the field $\phi(x,z)$, but in terms of its associated quantum particle, summing over all particle trajectories, $X^m(s)\equiv\{x^\mu(s),z(s)\}$, that connect the two desired points $x^m_1$ and $x^m_2$ in the bulk spacetime: 
\begin{equation}
\label{eqn:geodesic}
 \expval{\phi(x^m_1)\phi(x^m_2)}_{\bulk}=
 \int_{X(0)=x_1}^{X(1)=x_2}
 \!\!\!\mathcal{D}X^{m}(s)\,e^{i \ibulk[X]}~. 
\end{equation}
 The particle action is $\ibulk=M\mathsf{T}[X]$, where $\mathsf{T}$ denotes the proper time of the trajectory.\footnote{Details on the path integral measure can be found, e.g., in \cite{louko}, but we will not need them here, since we will ultimately work only in the saddle-point approximation.} Spacelike paths are assigned positive imaginary proper time, $\mathsf{T}=i\,\mathsf{L}$, with $\mathsf{L}$ the proper length. 
We can then  insert (\ref{eqn:geodesic}) in (\ref{eqn:BDHM}) and, for large $ML\simeq\Delta$, use the saddle point approximation
\cite{balasubramanianross,louko}. As we will see below, a short calculation shows that, in pure AdS, this method correctly reproduces the functional form of the CFT 2-point function, which is fixed by conformal symmetry.

It will be convenient for us to fix the static gauge
$s=z$, describing the paths as $X^{\mu}(z)$ (with an implicit decomposition to account for the double- or multi-valuedness).
We also introduce the usual UV cutoff at $z=\eps$, to regularize the divergent boundary-to-boundary length $\mathsf{L}$. We could add the appropriate counterterm $\iuv$ to eliminate this divergence, but as in the preceding sections, it plays no role in the rememorization procedure, so we will omit it.  The path integral in (\ref{eqn:geodesic}) then takes the form
\begin{equation}
\label{eqn:path02}
 \expval{\phi(x_1,\epsilon)\phi(x_2,\epsilon)}_{\bulk}=
 \int_{X(\epsilon)=x_1}^{X(\epsilon)=x_2}
 \!\!\!\mathcal{D}X^{\mu}(z) e^{i\ibulk[X]}~,
\end{equation}
with
\begin{equation}
\label{ibulkgeodesic}
    i\ibulk[X]=
    -ML\int_{\epsilon}^{\bar{z}}dz \frac{\sqrt{\dot{X}\cdot\dot{X} +1}}{z}~,
\end{equation}
where $\cdot$ refers to contraction with the Minkowski metric. The associated equation of motion is of course the geodesic equation, and for use below we note that it stipulates that the conjugate momentum
\begin{equation}\label{piconstant}
    \Pi_{\mu}\equiv\frac{\dot{X}_{\mu}}{z\sqrt{\dot{X}\cdot\dot{X}+1}}=\mbox{constant}~.
\end{equation}

Now, given a choice of bulk region $\cR$ that intersects the boundary on some region $R$, rememorization will allow us to correctly compute, purely from within $\cR$, CFT correlators with an arbitrary number $n$ of insertions of $\cO$ inside $R$. Our task is thus to determine the IR boundary action $\iir$ that accounts for all possible portions of paths that lie in $\cR\sc$.

In the free limit, non-vanishing $n$-point correlators are products of propagators (\ref{eqn:path02}) with $n\in 2\mathbf{N}$, for all possible pairings of the insertion points $x_1,\ldots,x_n$. Enumerate the corresponding paths as
$X_{I}^{\mu}(z)$, with $I=1,\ldots,n/2$. For each $I$, we are considering paths that join a specific pair $(x_i,x_j)$ of the insertion points. Such paths can be 
divided into classes according to the number
$k_I$ of times that they cross $\rir$, the interface between $\cR$ and $\cR\sc$. The 
 path integral thus breaks up into a sum, 
$\cD X_I^{\mu}=\sum_{k_I\in 2\mathbf{N}}\cD X^{\mu}_{(k_I)}$. 

The total number of $\rir$ crossings of the $n/2$ paths is $k\equiv k_1+\ldots+k_{n/2}$. 
Let $\bar{x}_1,\bar{x}_2,\ldots,\bar{x}_k$ 
denote the location of the crossing points, with the corresponding radial positions $\bar{z}_1,\ldots,\bar{z}_k$ determined then by the embedding function of $\rir$, $\bar{z}(x)$.
For each value of $k$, the portions of the paths $X^{\mu}_{I}$ contained within $\cR\sc$ are  $k/2$ segments connecting the $k$ points $\bar{x}_i$ pairwise (and forbidden from crossing $\rir$ at other points). The segment connecting $\bar{x}_i$ and $\bar{x}_j$ will be referred to as  $X^{\mu}_{i,j}(z)$.  For each sector labeled by $k$, we define the IR boundary action as
\begin{equation}\label{iirkcrossings}
\exp\left(i\iir^{(k)}(\bar{x}_1,\ldots,\bar{x}_k)\right)
\equiv
\sum_{p}\int \prod_{<i,j>}\cD X^{\mu}_{i,j}(z)
     \exp\left(i(\ibulk^{\cR\sc}[X_{i,j}])\right)~,
\end{equation}
where the sum runs over the distinct pairings $p$ of the $\bar{x}_i$, and we refrain from introducing additional notation to indicate the explicit manner in which the pairs $<\!\!i,j\!\!>$ are determined by $p$. The  full boundary action $\iir$ receives contributions from $\iir^{(k)}$ for all values of $k$. 

The main point to appreciate here is that, in parallel with (\ref{iraction}), the EOW brane action \iir\ is defined without any reference to $\cR$, and is therefore agnostic about the number $n$ and locations $x_i$ of the operators that one might later wish to insert inside $R$.\footnote{As explained in the Introduction, to resolve the puzzle encountered in \cite{bao}, we would need to rememorize string worldsheets instead of geodesics. In that case, the boundary action $\iir$ would be again agnostic about the number and location of the corresponding Wilson loops that are to be inserted later within $R$, so it would necessarily be a functional of all possible curves on which such worldsheets can intersect $\rir$.} 
Having emphasized this point, we will focus henceforth on the case $n=2$ that is the building block for all higher-point correlators. The two insertion points will be labeled $x$ and $y$. And, anticipating our use of the saddle-point approximation, we will work out (\ref{iirkcrossings}) just in the case $k=2$, which will yield the only non-vanishing saddle. The two $\rir$ crossing points will be denoted $\bar{x}$ and $\bar{y}$. The definition (\ref{iirkcrossings}) then simplifies to
\begin{equation}\label{iir2crossings}
\exp\left(i\iir^{(2)}(\bar{x},\bar{y})\right)
\equiv
\int \cD X^{\mu}_{\bar{x},\bar{y}}(z)
     \exp\left(i(\ibulk^{\cR\sc}[X_{\bar{x},\bar{y}}])\right)~,
\end{equation}
and the path integral (\ref{eqn:path02}) is rewritten in rememorized form as
\begin{eqnarray}
\label{rememorizedpaths}
 \expval{\cO(x)\cO(y)}_{\cft}&\propto&
 \int d\bar{x}d\bar{y}\int
 \mathcal{D}X^{\mu}_{x,\bar{x}}(z)
 \mathcal{D}X^{\mu}_{y,\bar{y}}(z)\\
 {}&{}&\qquad\qquad
 \exp\left(i(\ibulk^{\cR}[X_{x,\bar{x}}]
 +\ibulk^{\cR}[X_{y,\bar{y}}]
 +\iir^{(2)}(\bar{x},\bar{y}))\right)~.\nonumber
\end{eqnarray}
It is understood here that the unbarred points $x,y$ lie on the UV cutoff surface $z=\eps$, while the barred points $\bar{x},\bar{y}$ are on the EOW brane at $\rir$, at the appropriate radial position $z=\bar{z}(\bar{x})$ or $\bar{z}(\bar{y})$. For large $\Delta$, the path integrals are 
well-approximated by the corresponding saddles.

In the remainder of the present section, we will apply this procedure to a couple of examples of interest. 
For simplicity, we will work with a bulk region $\cR$ in  Poincar\'e-AdS$_{3}$, but the saddles we encounter would also be relevant in situations with appropriate symmetry in higher-dimensional pure AdS, and the same rememorization procedure would work as well for arbitrary bulk regions in general asymptotically locally AdS$_{d+1}$ settings. 

The original computation (\ref{eqn:path02}), without any splitting, yields
\begin{equation} \label{fullsaddle}
 \expval{\cO(x)\cO(y)}_{\cft}\propto\,
 \exp(i\ibulk[X^{\text{cl}}_{x,y}])~,
\end{equation}
with $X^{\text{cl}}_{x,y}$ the geodesic (a solution of (\ref{piconstant})) connecting $x$ and $y$,\footnote{Here we choose to enforce the boundary conditions on the geodesic at $z=0$ instead of $z=\epsilon$, which is equally valid, since the difference between the two choices is subleading in $\epsilon$.} 
\begin{equation} \label{fullgeodesic}
\left(x-X^{\text{cl}}_{x,y}\right)\cdot
\left(X^{\text{cl}}_{x,y}-y\right)=z^2~.
\end{equation}
This is a (generally boosted) semicircle centered on the AdS boundary \cite{rt,myers},
with regularized proper length
\begin{equation}\label{fulllength}
    \mathsf{L}[X^{\text{cl}}_{x,y}])= 2L\ln\!\left(\frac{\norm{x-y}}{\epsilon}\right)~,
\end{equation}
where $\norm{\cdot}$ denotes the norm with respect to the Minkowski metric. Recalling that $i\ibulk=-M\mathsf{L}$ and $ML\simeq\Delta$, (\ref{fulllength}) implies
\begin{equation}\label{fullibulk}
    i\ibulk[X^{\text{cl}}_{x,y}]=-2\Delta\ln\!\left(\frac{\norm{x-y}}{\epsilon}\right)~.
\end{equation}
Upon substituting this in (\ref{fullsaddle}), we recover the behavior
$\expval{\cO(x)\cO(y)}_{\cft}\propto\,\norm{x-y}^{-2\Delta}$ demanded by conformal invariance. Equation~(\ref{fullibulk}) is then what the rememorized path integral (\ref{rememorizedpaths}) must reproduce.

\begin{figure}[!t]
\begin{center}
\includegraphics[width=7cm,trim={1mm 1mm 1mm 1mm},clip]{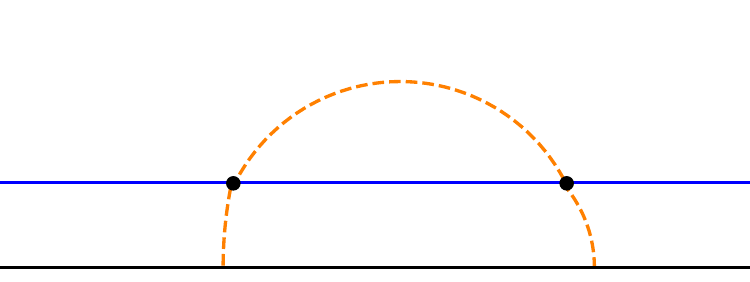}\hspace{10mm}\includegraphics[width=7cm,trim={1mm 1mm 1mm 1mm},clip]{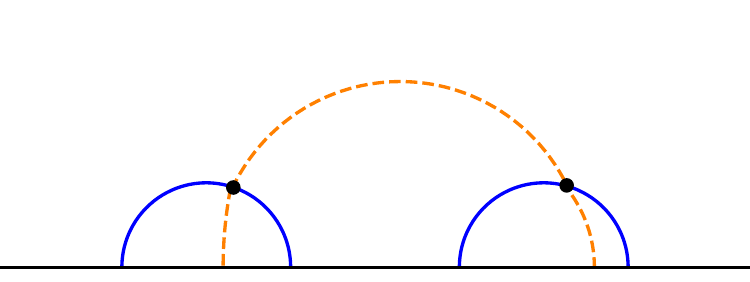}
  \hspace*{2cm}
  \setlength{\unitlength}{1cm}
\begin{picture}(0,0)
\put(4.8,1.6){$\bar{y}$}
\put(1.3,1.6){$\bar{x}$}
\put(4.8,0.2){$y$}
\put(1.3,0.2){$x$}
\put(-3.2,1.6){$\bar{y}$}
\put(-6.7,1.6){$\bar{x}$}
\put(-3.2,0.2){$y$}
\put(-6.7,0.2){$x$}
\end{picture}
\end{center}
\vspace*{-0.8cm}
\caption{When rememorizing to a bulk region $\cR$, geodesics that compute a two-point function $\expec{\cO(x)\cO(y)}$ can exit the region of interest. The portion of the geodesic in $\cR\sc$ will then be encoded in the EOW brane action $\iir$. The figure on the left shows the example analyzed in \S~\ref{wallgeodesicsubsec}, where the interface $\rir$ (solid blue) is a wall at fixed radial depth. On the right, we see the example of \S~\ref{doubleintervalsubsec}, where $\rir$ is the disconnected RT surface associated with two distant intervals.  
\label{geodesicrememorization}}
\end{figure}

In Section \ref{wallgeodesicsubsec}, we first examine the setup shown in  Fig.~\ref{geodesicrememorization} left, in which we place a wall at constant $z=\zir$ in  Poincar\'e-AdS$_{3}$, and  take $\cR$ to be the UV region $z\le\zir$. By symmetry, this analysis is relevant also for the higher-dimensional entanglement wedge $\cE$ of a strip, shown (at fixed time) in Fig.~\ref{geoexamplesfig} right, as long as the operators are inserted at the same time $t$ (not seen in the figure) and the same value of the coordinate along the width of the strip ($x_2$ in the figure).

As a second example, in \S~\ref{doubleintervalsubsec} we integrate out the region exterior to the entanglement wedge $\cE$ of two disjoint intervals in pure AdS$_3$. For convenience, we pick a configuration where the two intervals are sufficiently far away, in such a way that the entanglement wedge is in the disconnected configuration. See Fig.~\ref{geodesicrememorization} right. Conceptually, the saddle-point calculation in this setup is the same as the one relevant for the annulus in Fig.~\ref{geoexamplesfig} left, with insertions at diametrically opposite points, although the cross section of the annulus does not yield precisely semicircles. It is also worth noting that the type of EOW brane action that we obtain for this example, which couples the two RT surfaces nonlocally, is qualitatively the same as would be expected in the case of entanglement islands \cite{penington,aemm,islands}.

In each case, we will show that rememorization  produces an EOW brane action $\iir$ that encodes the relevant information about the part of the geometry that is integrated out, thus preserving the memory of the original global state. As emphasized before, this memory is not perfect, in the sense that there can be multiple global states that yield the same IR boundary action. Geodesics in the presence of EOW branes have also been considered in recent computations of correlators in AdS/BCFT \cite{adsbcftnotgeneric,shashi}, but the boundary action there is meant to encode a boundary condition for the field $\phi$ that is different from what we seek here.

\subsection{Wall at constant $z$} \label{wallgeodesicsubsec}

In the first setup, we consider a wall at $z=\bar{z}$ in AdS$_3$, and integrate out the region in the IR, $z>\bar{z}$, in the saddle-point approximation. What remains, the UV region $z\le\bar{z}$, is then our choice of $\cR$.\footnote{As explained three paragraphs above, this setup is also relevant for the entanglement wedge of a strip in the higher-dimensional situation shown in Fig.~\ref{geoexamplesfig} right.} The presence of the wall implies that there are two types of correlators, depending on whether $\norm{x-y}$ is smaller or larger than $2\bar{z}$. The first type is somehow trivial, given in terms of a geodesic (\ref{fullgeodesic}) that fits completely in the UV portion of the geometry. Since the purpose of our analysis is to show how the presence of the boundary term encodes information of the IR, we will not consider this case here (in the language of (\ref{iirkcrossings}),
it corresponds to the $k=0$ saddle, $\iir^{(0)}=0$).

The second type of correlator, with $\norm{x-y}>2z$, is given in terms of a geodesic (\ref{fullgeodesic}) that penetrates beyond the wall, thus probing $\cR\sc$, the IR portion of the geometry. See Fig.~\ref{geodesicrememorization} left. In this case, the saddle result for the boundary term (\ref{iir2crossings}) is
$-M\mathsf{L}
[X^{\text{cl}}_{\bar{x},\bar{y}}]$. A brief calculation yields the explicit expression
\begin{equation}
\label{eqn:lir}
 i\iir^{(2)}(\bar{x},\bar{y}) =-2\Delta\ln( \frac{\norm{\bar{x}-\bar{y}}
+\sqrt{\norm{\bar{x}-\bar{y}}^2 +4\bar{z}^2}}{2\bar{z}})~.
\end{equation}
Used in (\ref{rememorizedpaths}), this gives a complete specification of the problem, purely within $\cR$, in terms of the total effective action
\begin{equation}\label{itot}
I\eff\equiv\ibulk^{\cR}[X_{x,\bar{x}}]
 +\ibulk^{\cR}[X_{y,\bar{y}}]
 +\iir^{(2)}(\bar{x},\bar{y})~.
\end{equation}

Let us now see whether (\ref{eqn:lir}) does the job that it is supposed to. As explained in Section~\ref{dnsubsec}, there are two perspectives we can take on this, depending on whether we choose to employ the Dirichlet or Neumann boundary conditions on $\rir$ allowed by the variational principle based on 
$I\eff$. 

\subsubsection*{Dirichlet approach}

The most direct possibility is to work with the Dirichlet choice of boundary conditions on $\rir$.  This is evidently what is called for within the integrand of the $d\bar{x}$, $d\bar{y}$ integrals. We thus need to take $\bar{x},\bar{y}$ fixed and arbitrary, find the two UV saddles associated with $\ibulk^{\cR}[X_{x,\bar{x}}]$ and $\ibulk^{\cR}[X_{y,\bar{y}}]$, and carry out the integral over $\bar{x}$ and $\bar{y}$ at the end. 

The required saddles again involve $M\mathsf{L}$, with $\mathsf{L}$ the proper length of the relevant geodesics. Explicit computation yields
\begin{equation}
\label{eqn:luv1}
   i\ibulk^{\cR}[X^{\text{cl}}_{x,\bar{x}}] =-\Delta\ln(\frac{\norm{x-\bar{x}}^2 +\bar{z}^{2}}{\epsilon \bar{z}})~,
\end{equation}
and then of course
\begin{equation}
\label{eqn:luv2}
    i\ibulk^{\cR}[X^{\text{cl}}_{y,\bar{y}}] =-\Delta\ln(\frac{\norm{y-\bar{y}}^2 +\bar{z}^{2}}{\epsilon \bar{z}})~.
\end{equation}

To complete the calculation of (\ref{rememorizedpaths}), we are required to perform the remaining integrals, which run over the arbitrary points $\tilde{x}$ and $\tilde{y}$. Once again, we make use of the saddle point approximation. To obtain the on-shell final values for $\bar{x}$ and $\bar{y}$, we extremize $I\eff$ with respect to these two variables. This results in the following conditions:
%%double-check
\begin{align}
  \frac{ ( x_{\mu}-\bar{x}_{\mu})}{\norm{x-\bar{x}}^2 +\bar{z}^2} -\frac{( \bar{x}_{\mu}-\bar{y}_{\mu})}{\norm{\bar{x}-\bar{y}}\sqrt{\norm{\bar{x}-\bar{y}}^2 +4\bar{z}^2}}=0\,,\\
  \frac{( y_{\mu}-\bar{y}_{\mu})}{\norm{y-\bar{y}}^2 +\bar{z}^2}+\frac{( \bar{x}_{\mu}-\bar{y}_{\mu})}{\norm{\bar{x}-\bar{y}}\sqrt{\norm{\bar{x}-\bar{y}}^2 +4\bar{z}^2}}=0\,.
\end{align}
These conditions force $\bar{x}$ and $\bar{y}$ to lie, as expected, along the line that connects $x$ and $y$. Letting $u$ be the unit vector in this direction, $u\equiv (x-y)/\norm{x-y}$, the system of equations admits the solution
\begin{align}
\label{eqn:sol01}
     x^{\mu}-\bar{x}^{\mu} &=u^{\mu}( l-\sqrt{l^2-\bar{z}^2})~,\\
     \label{eqn:sol02}
     \bar{x}^{\mu}-\bar{y}^{\mu} &=2 u^{\mu}\sqrt{l^2 -\bar{z}^2}~,\\
     \label{eqn:sol03}
     y^{\mu}-\bar{y}^{\mu} &=-u^{\mu}( l-\sqrt{l^2-\bar{z}^2})~,
\end{align}
where  $l$ is short for $\norm{x-y}/2$. Substitution of (\ref{eqn:sol01})-(\ref{eqn:sol03}) into (\ref{eqn:lir}), (\ref{eqn:luv1})  and (\ref{eqn:luv2}) gives the following value for the action (\ref{itot}) at the saddle:
\begin{align}
\label{eqn:valact}
    iI\eff^{\mbox{\tiny{on-shell}}}
    =-2\Delta\ln\!\left(\frac{\norm{x-y}}{\epsilon}\right)~.
\end{align}
This matches the expected result (\ref{fullibulk}). 

\subsubsection*{Neumann approach}
An equivalent, but shorter route is to take into account from the beginning that the net effect of the $\bar{x}$ and $\bar{y}$ integrals is to enforce the Neumann boundary condition associated with $I\eff$ (which in turn amounts to the statement that the slopes of the IR and UV geodesics must agree). We can then dispense with these integrals, and simply look at once for the  Neumann saddle, which by construction should match the saddle of the full problem.  

Starting from the effective action
\begin{align}
\label{eqn:Sneu}
    iI\eff&=
    -\Delta\int_{\epsilon}^{\bar{z}}dz\left\{ \frac{\sqrt{\dot{X}_{x,\bar{x}}\cdot\dot{X}_{x,\bar{x}} +1}}{z}
    +\frac{\sqrt{\dot{X}_{y,\bar{y}}\cdot\dot{X}_{y,\bar{y}} +1}}{z}\right\}
    \nonumber\\
    {}&\qquad\qquad\qquad
    -2\Delta\ln\!\left( \frac{\norm{\bar{x}-\bar{y}}+\sqrt{\norm{\bar{x}-\bar{y}}^2 +4\bar{z}^2}}{2\bar{z}}\right)~,
\end{align}
extremization with arbitrary $\delta\bar{x}$ and $\delta\bar{y}$ yields, aside from the conservation equation (\ref{piconstant})
for both $X^{\mu}_{x,\bar{x}}(z)$ and $X^{\mu}_{y,\bar{y}}(z)$, the Neumann boundary conditions
%%double-check
\begin{align} \label{xbc}
  \Pi^{\mu}_{x,\bar{x}}|_{z=\bar{z}}&=
  -\frac{2(\bar{x}^{\mu}-\bar{y}^{\mu})}{\norm{\bar{x}-\bar{y}}\sqrt{\norm{\bar{x}-\bar{y}}^2 +4\bar{z}^2}}~,\\
  \label{ybc}
  \Pi^{\mu}_{y,\bar{y}}|_{z=\bar{z}}&=
  +\frac{2(\bar{x}^{\mu}-\bar{y}^{\mu})}{\norm{\bar{x}-\bar{y}}\sqrt{\norm{\bar{x}-\bar{y}}^2 +4\bar{z}^2}}~.
  \end{align}
The system of equations (\ref{piconstant}), (\ref{xbc}), (\ref{ybc}) leads correctly to the known solution (\ref{eqn:sol01})-(\ref{eqn:sol03}), as can be checked by simple substitution. 
%%these 3 eqns are just the locations of the correct xbar,ybar. It's a bit misleading to say here that we have solved the geodesiceom. But the general solution is known to be the semicircle given above.
This again guarantees that (\ref{eqn:valact}) is fulfilled, thereby recovering (\ref{fullibulk}).
We thus see that the Neumann approach indeed constitutes a useful shortcut, where we land immediately on the desired saddle. 

\subsection{Double interval} \label{doubleintervalsubsec}

For our second example, we consider two intervals of equal width $2l$ on a common time slice on the boundary of Poincar\'e AdS$_3$, with distance $2d$ between their centers.\footnote{So $d$ in this subsection should not be confused with the arbitrary boundary spacetime dimension referred to in the rest of the paper, here fixed at 2.} This arrangement of intervals, depending on the separation $2d$, has two possible configurations for its entanglement wedge $\cE$ \cite{headrick}. The first corresponds to the case where the separation between intervals is small ($d<l/\sqrt{2}$), and gives rise to a connected entanglement wedge, whereas the second configuration refers to large separations ($d>l/\sqrt{2}$) and results in a disconnected entanglement wedge $\cE$. For our purposes, this second case is more interesting, because a geodesic running from one interval to the other necessarily exits $\cE$.\footnote{As is well known, in the disconnected case the mutual information between the two intervals is zero only at leading order in $1/c$, and it is the subleading piece that allows the 2-point function under consideration to be nonvanishing \cite{lm,af}.}   See Fig.~\ref{geodesicrememorization} right.\footnote{As explained two paragraphs below (\ref{fullibulk}), this setup is conceptually similar to the entanglement wedge of an annulus, in the higher-dimensional situation shown in Fig.~\ref{geoexamplesfig} left.} The analysis runs in parallel with that of the previous subsection, so we will be brief. 

The boundary action on $\eir$ is found to be
\begin{align}
\label{eqn:lir02}
 i\iir^{(2)}(\bar{x},\bar{y}) &= -\Delta\ln\!\left( \frac{\sqrt{l^2-(\bar{x}-d)^2}}{\bar{x}(\bar{y}-\bar{x})+(\bar{y}+\bar{x})d+\sqrt{l^2 (\bar{x}-\bar{y})^2 +4d\bar{x}\bar{y}(d+\bar{y}-\bar{x})}}\right)
 \\
 & \quad
 +\Delta\ln\!\left( \frac{\sqrt{l^2-(\bar{y}+d)^2}}{\bar{y}(\bar{y}-\bar{x})+(\bar{y}+\bar{x})d+\sqrt{l^2 (\bar{x}-\bar{y})^2 +4d\bar{x}\bar{y}(d+\bar{y}-\bar{x})}}\right)~.
 \nonumber
 \end{align}
 Use of this in (\ref{rememorizedpaths}) gives a complete formulation of our problem purely within $\cR=\cE$.
 
 \subsubsection*{Dirichlet approach}
 Considering $\bar{x}$ and $\bar{y}$ to be fixed and arbitrary, we obtain the following two saddles within $\cE$:
\begin{align}
 \label{eqn:luv0202}
     i\ibulk^{\cE}[X^{\text{cl}}_{x,\bar{x}}] 
     &=-\Delta\ln\! \left(\frac{l^2+(x-d)(x+d-2\bar{x})}{\epsilon\sqrt{l^2 -(\bar{x}-d)^2}} \right)~,\\
 \label{eqn:luv0203}     
    i\ibulk^{\cE}[X^{\text{cl}}_{y,\bar{y}}]
    &=-\Delta\ln\! \left(\frac{l^2-(y+d)(y-d-2\bar{y})}{\epsilon\sqrt{l^2 -(\bar{y}+d)^2}} \right)~.
\end{align}

Next, we extremize the effective action (\ref{itot}) with respect to $\bar{x}$ and $\bar{y}$. The resulting conditions  are
%%double-check
\begin{align}
 &\frac{l^2 (\bar{x}-\bar{y})-2d(\bar{x}-d)\bar{y}}{\left[l^2 -(\bar{x}-d)^2 \right]\sqrt{l^2 (\bar{x}-\bar{y})^2 +4d\bar{x}\bar{y}(d+\bar{y}-\bar{x})} }   \nonumber\\
& \qquad \qquad \quad +\frac{1}{2(l+d-\bar{x})}-\frac{1}{2(l-d+\bar{x})}
 -\frac{2(x-d)}{l^2 +x^2-d^2 -2(x-d)\bar{x}}=0~,\\
 &\frac{l^2 (\bar{x}-\bar{y})-2d(\bar{y}+d)\bar{x}}{\left[l^2 -(\bar{y}+d)^2 \right]\sqrt{l^2 (\bar{x}-\bar{y})^2 +4d\bar{x}\bar{y}(d+\bar{y}-\bar{x})} } \nonumber \\
& \qquad \qquad \quad +\frac{1}{2(l-d-\bar{y})}-\frac{1}{2(l+d+\bar{y})}-\frac{2(y-d)}{l^2 +y^2-d^2 -2(y+d)\bar{y}}=0~.
\end{align}
The solution is
\begin{align}
\label{eqn:solInterval01}
    \bar{x}=\frac{l^2 -d^2 +xy}{x+y-2d}~,\\
\label{eqn:solInterval02}
    \bar{y}=\frac{l^2 -d^2 +xy}{x+y+2d}~.
\end{align}
If we substitute this into (\ref{eqn:lir02})-(\ref{eqn:luv0203}), we find again (\ref{eqn:valact}), meaning that we correctly recover the known result (\ref{fullibulk}).

\subsubsection*{Neumann approach}

The main novelty with respect to the wall case of the previous subsection is that the bulk action (\ref{ibulkgeodesic}) now includes an $\bar{x}$-dependent limit in the integral $\int_{\epsilon}^{\bar{z}(\bar{x})}dz$ (and similarly for $\bar{y}$), which contributes an additional term
when we extremize the effective action (\ref{itot}) keeping $\delta\bar{x}$ and $\delta\bar{y}$ arbitrary. The resulting Neumann boundary conditions read
%%double-check
\begin{align}
    \eval{\frac{\sqrt{1+\dot{X}^2_{x,\bar{x}} }}{z}}_{\bar{z}(\bar{x})}\frac{d-\bar{x}}{\sqrt{l^2 -(\bar{x}-d)^2}}
    %+\eval{\frac{\dot{X}_{x,\bar{x}}}{z\sqrt{1+\dot{X}^2_{x,\bar{x}}}}
    +\eval{\Pi_{x,\bar{x}}}_{\bar{z}(\bar{x})}
    -\frac{i}{\Delta}\frac{\partial \iir^{(2)}}{\partial \bar{x}}=0~,\\
     -\eval{\frac{\sqrt{1+\dot{X}^2_{y,\bar{y}} }}{z}}_{\bar{z}(\bar{y})}\frac{d+\bar{y}}{\sqrt{l^2 -(\bar{y}+d)^2}}
     %+\eval{\frac{\dot{X}_{y,\bar{y}}}{z\sqrt{1+\dot{X}^2_{y,\bar{y}}}}
     +\eval{\Pi_{y,\bar{y}}}_{\bar{z}(\bar{y})}
     -\frac{i}{\Delta}\frac{\partial \iir^{(2)}}{\partial \bar{y}}=0~.
\end{align}
%%explain meaning of dx/dxbar (through inverse)
Calculation reveals that (\ref{eqn:solInterval01}) and (\ref{eqn:solInterval02}) are solutions to these conditions, so once again, the Neumann approach proves to be a shortcut that correctly leads us back to (\ref{fullibulk}).  

\section*{Acknowledgements}

We are grateful to Daniel \'Avila, Ning Bao, Bartek Czech, Hao Geng, Matt Headrick, Jonathan Heckman, Ayan Mukhopadhyay, Mart\'in Sasieta and Zixia Wei for comments on the first version of this paper. 
The work of AG and YDO is partially supported by Mexico's National Council of Science and Technology (CONACyT) grant A1-S-22886 and DGAPA-UNAM grant IN107520. JFP is supported by ``la Caixa'' Foundation (ID 100010434), fellowship code LCF/BQ/PI21/11830029, and by the European Union's Horizon 2020 research and innovation programme under the Marie Sk{\l}odowska-Curie grant agreement No.~847648.

%%arXiv.org - Non-exclusive license to distribute

%%remove unused refs below %-

\end{document}